\documentclass[aps,prd,groupedaddress,11pt]{revtex4-2}

\usepackage{graphicx}
\usepackage{dcolumn}
\usepackage{bm}
\usepackage{amssymb}
\usepackage{amsmath}
\usepackage{epsfig}    
\usepackage{color}
\usepackage{slashed}
\usepackage{hhline}
\usepackage{placeins}

\def\be{\begin{equation}}
\def\ee{\end{equation}}
\newcommand{\bea}{\begin{align}}
\newcommand{\eea}{\end{align}}
\newcommand{\nn}{\nonumber}

\newcommand{\secref}[1]{Section~\ref{#1}}

\def\beq{\begin{equation}}
\def\eeq{\end{equation}}
\def\bma{\begin{math}}
\def\ema{\; \end{math}}
\def\begary{\beq\begin{array}{c}}
\def\endary{\end{array}\eeq}
\def\begaryl{\beq\begin{array}{l}}
\def\endaryl{\end{array}\eeq}
\def\bmat{\left( \begin{array}{ccc}}
\def\emat{\end{array}\right)}
\def\bmatv{\left( \begin{array}{cccc}}
\def\ematv{\end{array}\right)}
\def\bmatz{\left( \begin{array}{cc}}
\def\ematz{\end{array}\right)}
\def\bvec{\left( \begin{array}{c}}
\def\evec{\end{array}\right)}

\def\X2{$\chi^2$}
\def\EP{$e^{-}+e^{+}$}

\numberwithin{equation}{section}

\begin{document}

\preprint{APCTP Pre2020 - 006}

\title{Cosmic-Ray Signatures of Dark Matter from a Flavor Dependent Gauge Symmetry Model with Neutrino Mass Mechanism} 
%
\author{Holger Motz}
\email{motz@aoni.waseda.jp}
\affiliation{Global Center for Science and Engineering, Faculty of Science and Engineering, Waseda University, Shinjuku 169-8555, Tokyo, Japan}

\author{Hiroshi Okada}
\email{hiroshi.okada@apctp.org}
\affiliation{Asia Pacific Center for Theoretical Physics (APCTP) - Headquarters San 31, Hyoja-dong,
Nam-gu, Pohang 790-784, Korea}
\affiliation{Department of Physics, Pohang University of Science and Technology, Pohang 37673, Republic of Korea}

\author{Yoichi Asaoka}
\email{yoichi.asaoka@aoni.waseda.jp}
\affiliation{Waseda Research Institute for Science and Engineering, Waseda University, Shinjuku 169-8555, Tokyo, Japan}

\author{Kazunori Kohri}
\email{kohri@post.kek.jp}
\affiliation{The Graduate University of Advanced Studies (Sokendai),
Tsukuba 305-0801, Japan}
\affiliation{Theory Center, IPNS, KEK, Tsukuba 305-0801, Japan}
\affiliation{Kavli IPMU (WPI), UTIAS, The University of Tokyo,Kashiwa, Chiba 277-8583, Japan}

\date{\today}

\begin{abstract}
We propose an extension to the Standard Model accommodating two families of Dirac neutral fermions and Majorana fermions under additional $U(1)_{e-\mu}\times Z_3\times Z_2$ symmetries where $U(1)_{e-\mu}$ is a flavor dependent gauge symmetry related to the first and second family of the lepton sector, which features a two-loop induced neutrino mass model. The two families are favored by minimally reproducing the current neutrino oscillation data and two mass difference squares and canceling the gauge anomalies at the same time. As a result, we have a prediction for neutrino masses.
The lightest Dirac neutral fermion is a dark matter candidate with tree-level interaction restricted to electron, muon and neutrinos, which makes it difficult to detect in direct dark matter search as well as indirect search focusing on the $\tau$-channel, such as through {$\gamma$-rays}. It may however be probed by search for dark matter signatures in electron and positron cosmic rays, and allows interpretation of a structure appearing in the CALET electron+positron spectrum around 350-400 GeV as its signature, with a boost factor $\sim$40 Breit-Wigner enhancement of the annihilation cross section. 
\end{abstract}
\maketitle
\newpage

\section{Introduction}
\label{intro}

The cosmological standard model includes dark matter (DM) as an essential component, commonly considered to be a neutral particle not part of the standard model of particle physics (SM). Assuming thermal production in the early Universe, a weakly interacting massive particle (WIMP) in the GeV-TeV mass is a strong candidate, since the Weak Interaction of the SM yields just the right annihilation cross section to predict the observed relic density of DM, a relation known as the WIMP miracle.
This default candidate is the main target of experimental DM search, and since the weak interaction couples universally to all leptons and quarks, its parameter space is successively scanned and ruled out by direct detection experiments based on WIMP-nucleon interactions~\cite{Escudero:2016gzx,Baer:2016ucr,Liu:2017kmx,Aprile:2018dbl,Aprile:2019dbj} 
and indirect searches looking for the products of annihilation into hadronic channels, such as anti-protons~\cite{Evoli:2015vaa,Cuoco:2017iax} and {$\gamma$-rays}~\cite{Abdalla:2018mve}. 

Avoiding hadronic interaction of DM requires the introduction of a new force and corresponding charge, which is only carried by the DM and leptons. In the initial version of this Leptophilic Dark Matter~\cite{Fox:2008kb}, 
all lepton generations carry the same charge, resulting in equal branching ratios in the annihilation of DM. In this case, the strongest constraints on the DM annihilation cross section come from observation of dwarf galaxies in {$\gamma$-rays} 
based on the $DM + DM \rightarrow \tau^{+} + \tau^{-}$ channel, which due to its higher {$\gamma$-ray} multiplicity yields limits about half a magnitude more strict than those on  $DM + DM \rightarrow e^{+} + e^{-}$ channel and $DM + DM \rightarrow \mu^{+} + \mu^{-}$ channel~\cite{Ackermann:2015zua,Archambault:2017wyh}. These limits are subject to about one order of magnitude variation from uncertainty on the halo shape and resulting J-factors~\cite{Hiroshima:2019wvj}, which however is independent of the annihilation channel.
Most recently very strict limits on hadronic and the $DM + DM \rightarrow \tau^{+} + \tau^{-}$ channel based on the morphology of {$\gamma$-ray} flux from the galactic center have been brought forward~\cite{Abazajian:2020tww}, giving explicitly no such constraint on $DM + DM \rightarrow e^{+} + e^{-}$ channel and $DM + DM \rightarrow \mu^{+} + \mu^{-}$ channel.

On the other hand, search for DM annihilation in positron and electron cosmic rays with detectors such as AMS-02~\cite{Aguilar:2013pos,Accardo:2014pos,Aguilar:2014ele,Aguilar:2014all,Aguilar:2019ele,Aguilar:2019owu}, CALET~\cite{Adriani:2017all,Adriani:2018ktz} , DAMPE~\cite{Ambrosi:2017all} and the Cosmic Ray Subsystem on the Voyager probes is most sensitive to the electron channel, since its signature is a sharp drop in the spectrum at the mass of the DM particle which can be recognized above a smooth astrophysical background~\cite{Ibarra:2013zia,Motz:2015cua,Boudaud:2016mos}. 
For GeV-TeV range DM, the target region is the local DM halo within $\sim$kpc range due to the energy loss and resulting limited propagation distance of electron cosmic rays.
This complementarity can reduce the possible impact of astrophysical uncertainties in the case of DM with universal coupling to leptons. For DM with selective coupling to the different lepton flavors, either search with {$\gamma$-rays} or charged cosmic rays may have preferential sensitivity. 

Apart from DM, the other strong indication of physics beyond the SM is the neutrino mass, and many theoretical models extending the SM aim at solving both issues simultaneously, examples being radiative seesaw models at one-loop~\cite{Ma:2006km}, two-loop~\cite{Kanemura:2011mw, Kajiyama:2013zla}, and three-loop~\cite{Krauss:2002px, Aoki:2008av, Gustafsson:2012vj}. 
Several models extending the SM by an additional  U(1) gauge symmetry have been proposed, which favor annihilation or decay to tau and/or muon as a possible DM-only explanation of the positron excess~\cite{Kohri:2013sva,Ko:2014lsa,Han:2019diw}, while also featuring a mechanism for giving the neutrinos mass. 

In this context we investigate if a thermally produced DM candidate based on a flavor-specific $U(1)_{e-\mu}$ gauge symmetry coupling only to electron and muon is also feasible, corroborated by simultaneous explanation of the neutrino sector. This kind of DM would be a favorable target to search in electron-positron cosmic rays while being less detectable by {$\gamma$-ray} search. After establishing the particle physics model 
defining the properties of the DM, we discuss its cosmic-ray signatures and implications from available CALET and AMS-02 data. 
While introduction of a new flavor-specific gauge interaction lacks the elegance of the classical WIMP, studying such a model seems worthwhile as it allows to keep a thermal production mechanism and a WIMP-like DM candidate. This should be seen against the trend of DM candidate theory becoming more and more diversified to avoid constraints on the WIMP and WIMP-like particles~\cite{Baer:2014eja}.

Our extension of the SM is based on a radiatively induced neutrino mass (scotogenic model), which originally provides us with an appropriate explanation of the hierarchy among the Yukawa sector of the SM. The ratio between the top Yukawa quark coupling($\sim1$) and the electron Yukawa coupling($\sim10^{-6}$) is of the order $10^6$, which respectively are the heaviest and lightest masses in the fermion sector of the SM.
However, the ratio between the electron Yukawa coupling and the typical neutrino Yukawa coupling($\sim10^{-13}$) is of the order $10^{7}$. If we assume the neutrino mass to be of Dirac type and to be induced at tree level, which is the same as for the other matter sectors in the SM, this would suggest that there is a huge gap 
between the neutrino coupling and the other three Yukawa couplings.
The scotogenic model generates neutrino mass at loop level, with newly introduced fields running inside the loop. It is found that with a 0.01 loop suppression factor and two Yukawa couplings at one-loop level in the neutrino mass formula, the order of Yukawa coupling at one-loop level is minimally $10^{-6}$, which is comparable to the electron Yukawa coupling. We fix the mass scale of one 
new field to be on the order of one TeV, which allows for the new scale to be tested by current experiments.
Another advantage of this model is its predicted correlation between the DM candidate properties and the neutrino mass, since the DM field is running inside the neutrino loop. Therefore, the neutrino interacts with SM-like Higgs only though the DM field in the generation of the neutrino mass. This provides a natural explanation for the tininess of the neutrino mass, and phenomenology apart from direct and indirect DM search.
%
Further phenomenology arises from the not so small strength of the Yukawa couplings and their nontrivial structure to induce the neutrino mixings as well as mass eigenvalues, which might cause lepton flavor violations (LFVs) that are severely restricted by current experiments such as MEG~\cite{TheMEG:2016wtm,Renga:2018fpd}. 

To realize a sufficiently high cross section yielding the observed relic density in thermal production of the DM candidate, the annihilation process should be s-channel dominated, which however is helicity-suppressed for a Majorana particle. Therefore we introduce a $Z_3$ discrete symmetry under which the DM is charged, giving it a Dirac nature and ensuring its stability.
Also, we impose a $Z_2$~discrete~symmetry to forbid tree level neutrino mass, where this symmetry is softly broken in the Higgs potential and its broken term contributes to generating the tiny neutrino mass.
The neutrino mass is induced at two-loop level, where we introduce two types of neutral fermions; Dirac type and Majorana type. 
In the neutrino sector, the $U(1)_{e-\mu}$ symmetry also plays an important role in predicting the neutrino mass.
Because the nonzero charges (-1 or 1) have to be assigned to only two families, the minimal number of new fermions are two families, which is also the minimal number to explain the active neutrino oscillation data and their mass eigenvalues.
Furthermore, the two families are required to allow gauge anomaly cancellation in a minimal manner.
Thus, we predict one massless neutrino that causes the other two massive neutrinos to be uniquely determined by the experimental results, which are the squared solar mass difference and squared atmospheric mass difference, as we will discuss for both cases of normal hierarchy (NH) and inverted hierarchy (IH) in detail.

This paper is organized as follows: 
In \secref{partphysmod}, we explain our particle physics scenario and formulate the lepton sector and the Higgs potential, the masses and mixings for the two new fermions and the active neutrinos, and the mass of the new gauge boson and its interactions, also discussing LFVs. 
In \secref{DMprop}, we will discuss our DM candidate, in which we briefly explain why it is not subject to current bounds from direct detection searches, and 
explain calculation of the DM relic density. 
We also show that Breit-Wigner enhancement may lead to a significant boost factor ($B$) on the annihilation cross section, which may increase the signatures to the level detectable by current indirect DM search. 
In \secref{DMsig}, the electron and positron cosmic-ray signature of the DM candidate X is explained, and after introduction of propagation and astrophysical background models, the \EP~flux measured by CALET~\cite{Adriani:2018ktz} and the $e^{+}$-only flux measured by AMS-02~\cite{Aguilar:2019owu} are interpreted including the DM signature. It is shown that step-like spectral structures in the CALET spectrum could be identified with the signature of the DM candidate, identifying the best-fit regions in $m_{\rm X}$ vs. $B$ space. 
Finally we summarize and conclude our results in \secref{sumconcl}.

 \section{Particle Physics Model}
\label{partphysmod}
In this section, we review our model~\cite{footnote1}.
At first, we explain our motivation for introducing new symmetries and fields.
Then, we construct the Lagrangian and Higgs potential, and continue with formulating the neutral fermions, LFVs, and the additional gauge boson sector.
 
\subsection{Particle Contents and Lagrangian}

We introduce three families of vector-like fermions ($N_{e},N_\mu,N_\tau$), and two families of
Majorana fermions ($\nu_{R_e},\nu_{R_\mu}$) in the fermion sector, so that
we can construct a two-loop induced neutrino mass model. These fermions are minimally required to reproduce the neutrino oscillation data and cancel the anomaly for $\nu_R$.
We extend the scalar sector by introducing an isospin doublet inert boson $\eta$, an isospin singlet inert boson $S$, and a singlet boson $\varphi$ that gives nonzero 
Vacuum Expectation Values (VEVs)
to spontaneously break the $U(1)_{e-\mu}$ symmetry as shown later, where the SM-like  scalar boson is symbolized by $H$. Here we denote their 
VEVs
as $\langle\varphi\rangle\equiv v_\varphi/\sqrt2$ and $\langle H\rangle\equiv v_H/\sqrt2$, respectively.
In addition, we impose three additional symmetries; gauged symmetry $U(1)_{e-\mu}$ and discrete Abelian symmetries $Z_3$ and $Z_2$.
The first symmetry defines the newly introduced interaction with only the two first generations of leptons, giving the model the intended property of avoiding gauge interactions with the $\tau$-lepton,
while the second one provides stability of potential DM candidates $N,\eta,S$, and assures the Dirac feature of $N$. We associate the lightest Dirac particle $N$ with DM, since the heavier ones can decay into the lighter ones via five-dimensional terms even though the decay is forbidden within the renormalizable theory. The field contents and their assignments are summarized in Table~\ref{tab:1} for fermions and Table~\ref{tab:2} for bosons.

\begin{table*}[t]
\begin{footnotesize}
\begin{tabular}{|c||c|c|c|c|c|c|c|c|c|c|c|c|}\hline
 & \multicolumn{6}{c|}{ SM fermions} & \multicolumn{5}{c|} {New fermions} \\\hline
Fermions  ~&~ $L_{L_e}$ ~&~ $L_{L_\mu}$ ~&~ $L_{L_\tau}$ ~&~ $e_R$ ~&~ $\mu_R$ ~&~ $\tau_R$ ~&~ $N_{e}$ ~&~ $N_{\mu}$ ~&~ $N_{\tau}$ ~&~ $\nu_{R_e}$ ~&~ $\nu_{R_\mu}$ ~
\\\hline 
$SU(3)_C$  & $\bm{1}$  & $\bm{1}$  & $\bm{1}$   & $\bm{1}$  & $\bm{1} $  & $\bm{1}$ & $\bm{1}$  & $\bm{1} $  & $\bm{1}$ & $\bm{1}$ & $\bm{1}$   \\\hline 
 $SU(2)_L$  & $\bm{2}$  & $\bm{2}$  & $\bm{2}$   & $\bm{1}$  & $\bm{1}$   & $\bm{1}$  & $\bm{1}$  & $\bm{1}$   & $\bm{1}$& $\bm{1}$& $\bm{1}$   \\\hline 
$U(1)_Y$  & $-\frac{1}{2}$ & $-\frac12$ & $-\frac12$  & $-1$ &  $-1$  &  $-1$  & $0$  & $0$ &  $0$  &  $0$   &  $0$ \\\hline
 $U(1)_{e-\mu}$ & $1$  & $-1$ & $0$ & $1$  & $-1$   & $0$ & $1$  & $-1$& $0$  & $1$  & $-1$  \\\hline
$Z_3$  & $1$  & $1$ & $1$ & $1$ & $1$ & $1$& $\omega$ & $\omega$& $\omega$ & $1$ & $1$ \\\hline
$Z_2$  & $+$  & $+$ & $+$ & $+$ & $+$ & $+$& $+$ & $+$ & $+$ & $-$ & $-$ \\\hline
\end{tabular}
\caption{Field contents of fermions
and their charge assignments under {$SU(2)_L\times U(1)_Y\times  U(1)_{e-\mu}\times Z_3\times Z_2$}, where $Z_2$ is softly broken and all the fields are singlet under $SU(3)_C$.\label{tab:1}}
 \end{footnotesize}
\vspace*{\floatsep}
\centering {\fontsize{10}{12}
\begin{tabular}{|c||c|c||c|c|c|}\hline
&\multicolumn{2}{c||}{VEV$\neq 0$} & \multicolumn{2}{c|}{Inert } \\\hline
  Bosons  &~ $H$  ~ &~ $\varphi$ ~ &~ $\eta$   ~ &~ $S$ ~ \\\hline
$SU(2)_L$ & $\bm{2}$ & $\bm{1}$ & $\bm{2}$  & $\bm{1}$    \\\hline 
$U(1)_Y$ & $\frac12$ & $0$ & $\frac12$  & $0$    \\\hline
 $U(1)_{e-\mu}$ & $0$  & $1$ & $0$ & $0$   \\\hline
$Z_3$ & $1$ & $1$  & $\omega$ & $\omega$ \\\hline
$Z_2$ & $+$ & $+$  & $+$ & $-$ \\\hline
\end{tabular}%
} 
\caption{Field contents of bosons
and their charge assignments under {$SU(2)_L\times U(1)_Y\times U(1)_{e-\mu}\times Z_3\times Z_2$}, where $SU(3)_C$ is singlet for all bosons, where $Z_2$ is softly broken, and all the fields are singlet under $SU(3)_C$. \label{tab:2}}
\end{table*}

{\it Anomaly cancellations}:
Since $U(1)_{e-\mu}$ gauge symmetry is anomaly free among the SM fermions, all we need to take into account is the new fermions. Furthermore, since all our fermions are neutral under $U(1)_Y$, we should consider the following two conditions: $U(1)_{e-\mu}$ and $[U(1)_{e-\mu}]^3$. Thus, one straightforwardly finds that these conditions are anomaly free in our model, since each of the fermions has opposite sign under the $U(1)_{e-\mu}$ charge.

{\it Yukawa Interactions}:
Under these fields and symmetries, the renormalizable Lagrangian for quark and lepton sector is given by 
\begin{align}
-{\cal L}_{L} = & \sum_{\ell=e,\mu,\tau} \sum_{\ell'=e,\mu}
 \left[y_{\ell} \bar L_{L_\ell}  H \ell_R+y_{\eta_\ell}\bar L_{L_\ell} \tilde\eta N_{R_\ell}
  + y_{S_{\ell'}} \bar N_{L_{\ell'}} \nu_{R_{\nu_{\ell'}}} S   +M_\ell \bar N_{R_\ell} N_{L_\ell} \right]
 \nn\\&
+\sum_{(\alpha,\beta)}^{(e,\tau),(\tau,\mu)}{f_\varphi}_{\alpha\beta}   \bar N_{R_\alpha} N_{L_\beta}\varphi 
+\sum_{(\alpha,\beta)}^{(\mu,\tau),(\tau,e)}{f'_\varphi}_{\alpha\beta}\bar  N_{R_\alpha} N_{L_\beta}\varphi^*  
+{M_N}_{e\mu} \bar \nu_{R_e} \nu^C_{R_\mu}
+ {\rm h.c.},
\label{eq:lag-quark}
\end{align}
where  $\tilde \eta$ is defined by $i\sigma_2 \eta^*$, $\sigma_2$ being the second Pauli matrix,
and $\bar N_R^C\nu_R S^*$ is also allowed by our symmetries but it does not contribute to any phenomenologies. Thus we neglect this term.
$Z_2$~symmetry forbids the Dirac term $\bar L_L\tilde H \nu_R$ at tree level, where $Z_2$ is softly broken at the Higgs potential below.

{\it Scalar potential}:
The renormalizable scalar potential is given by
\begin{align}
&V=-\mu_H^2 |H|^2 + m_\eta^2 |\eta|^2 - m_\varphi^2 |\varphi|^2 + \mu_S^2 |S|^2 \nn\\
&+(\mu H^\dag\eta S^* +{\rm h.c.}) + \lambda_{H}|H|^4 + \lambda_{\eta}|\eta|^4+ \lambda_{\varphi}|\varphi|^4
+ \lambda_{S}|S|^4 + \lambda_{H\eta}|H|^2|\eta|^2+ \lambda'_{H\eta}|H^\dag\eta|^2\nn\\
&+ \lambda'_{H\varphi}|H|^2|\varphi|^2+ \lambda'_{HS}|H|^2|S|^2 + \lambda_{\eta\varphi}|\eta|^2|\varphi|^2
+ \lambda_{\eta S}|\eta|^2|S|^2 + \lambda_{\varphi S}|\varphi|^2|S|^2,
\end{align}
where the $\mu H^\dag\eta S^*$ term is softly broken under $Z_2$ symmetry, and we expect $\mu$ to be of a rather small scale compared to the electroweak scale.
We parametrize the scalar fields as 
\begin{align}
&H =\left[
\begin{array}{c}
w^+\\
\frac{v_H+h+iz}{\sqrt2}
\end{array}\right],\ 
{\eta =\left[
\begin{array}{c}
\eta^+\\
{\eta_0}
\end{array}\right]} 
,\ 
\varphi=\frac{v'+\rho+i z'}{\sqrt2},\
\label{component}
\end{align}
where $\eta_0$ and $S$ are complex scalars,
$v_H~\simeq 246$ GeV is VEV of the SM Higgs, and $w^\pm$, $z$, and $z'$ are respectively Nambu-Goldstone bosons (NGB) 
which are absorbed by the longitudinal component of gauge bosons, denoted by $W$, $Z$, and $Z'$.
$Z'$ arises from the gauged $U(1)_{e-\mu}$ symmetry.
Then we have two neutral boson mass matrices $m^2_{h\rho}$,  $m^2_{\eta S}$ in the basis of $[h,\rho]^T$ and $[\eta_0,S]^T$, which are respectively diagonalized by $O_{a}^T m^2_{h\rho}O_{a}\equiv$Diag[$m_{h_1},m_{h_2}$] and   $O_{\alpha}^T m^2_R O_{\alpha}\equiv$Diag[$m_{H_1},m_{H_2}$],
where $m_{h_1}\approx125$ GeV is the mass of the SM Higgs.
Here we define the mixing matrices as
\begin{align}
O_{a(\alpha)} =
\left[\begin{array}{cc} c_{a(\alpha)} & s_{a(\alpha)} \\ -s_{a(\alpha)} & c_{a(\alpha)} \end{array}\right],\quad
s_{2a}=-\frac{2\lambda_{H\varphi} v v' }{m_{h_2}^2-m_{h_1}^2},\quad
s_{2\alpha}=- \frac{\sqrt2 \mu v_H }{m_{H_2}^2-m_{H_1}^2},
\label{eq:mixing}
\end{align}
where $c(s)_{a(\alpha)}$ is the short-hand notation of $\cos(\sin)_{a(\alpha)}$.
While values $s_a>0$ could be chosen within experimental limits, we take $s_a=0$ in our numerical analysis for convenience as shown later.

{\it Neutral Dirac Fermions}:
After the $e-\mu$ spontaneous breaking, the Dirac fermion mass matrix in basis of $[N_e,N_\mu,N_\tau]^T$ is found as:
\begin{align}
M_{N}\equiv \left[\begin{array}{ccc}  M_e &0 & M_{e\tau}  \\0 & M_\mu & M_{\mu\tau} \\ M_{\tau e} &  M_{\tau\mu} & M_\tau \end{array}\right]_{RL},\label{eq:ML'}
\end{align}
where $M_{e\tau}\equiv f_{\varphi_{e\tau}} v_\varphi/\sqrt2$, $M_{\tau\mu}\equiv f_{\varphi_{\tau\mu}} v_\varphi/\sqrt2$,
$M_{\mu\tau}\equiv f'_{\varphi_{\mu\tau}} v_\varphi/\sqrt2$, and $M_{\tau e}\equiv f'_{\varphi_{\tau e}} v_\varphi/\sqrt2$.
  $M_N$ is diagonalized by a bi-unitary mixing matrix as $D_N=V^\dag_R M_{N} V_L $:
\begin{align}
&V^\dag_R M_{N} M_N^\dag V_R =V^\dag_L M_{N}^\dag M_N V_L  \equiv {\text{Diag.} }\left[|M_1|^2,|M_2|^2,|M_3|^2\right],\\ &N_{L(R)_{e,\mu,\tau}}=V_{L(R)} \psi_{L(R)_{1,2,3}},\label{eq:N-mix}
\end{align}
where $M_{1,2,3}$ is the mass eigenstate, and $\psi$ is the mass eigenvector of $N$.

{\it Neutral Majorana Heavier Fermions}:
In a way similar to the Dirac fermion, the Majorana fermion mass matrix in the basis of $[\nu_{R_e},\nu_{R_\mu}]^T$ is found as:
\begin{align}
M_R \equiv \left[\begin{array}{cc}  0 & M_{N_{e\mu}}  \\ M_{N_{e\mu}}  & 0  \end{array}\right]_{}.\label{eq:MR'}
\end{align}
  $M_N$ is diagonalized by a unitary mixing matrix as $D_R=U^T M_{R}U$:
\begin{align}
&U^T M_{R} M_R^\dag U^*  \equiv {\text{Diag.} }\left[|M_{R_{1}} |^2,|M_{R_{2}}|^2\right],\\ &\nu_{R_{e,\mu}}= U \Psi_{R_{1,2}},\quad
U=\frac1{\sqrt2}\left[\begin{array}{cc}  1 & -1  \\1  & 1  \end{array}\right],
\label{eq:R-mix}
\end{align}
where $M_{R_{1,2}} = M_{N_{e\mu}}$ is the mass eigenstate, and $\Psi$ is the mass eigenvector  of $\nu_R$.

\subsection{Active Neutrino Mass}
The dominant contribution to the active neutrino mass matrix arises from the canonical seesaw model, but the Dirac mass matrix $m_D$ is given at one-loop level. Thus the neutrino mass is induced at two-loop level.
Before formulating the neutrino sector, we evaluate the number of complex parameters.
First of all, three components of $y_\eta$ can be real by phase redefinition for $L_{L_{e,\mu,\tau}}$, which implies that the phases of $N_{R_{e,\mu,\tau}}$ and $e_R,\mu_R,\tau_R$ are fixed.
Second, the two components of $y_S$ can also be real by the redefinition for $\nu_{R_{e,\mu}}$, which suggests that the phases of $N_{L_{e,\mu}}$ are fixed.
Finally, one phase in $M_N$ can be real by the phase redefinition for $N_{L_{\tau}}$. Here we identify $M_\tau$ to be real.
Thus, we have six phases in $M_N$.
The canonical seesaw is given by the following form:
\begin{align}
m_\nu\approx -m_D M_R^{-1} m_D^T,
\end{align}
where $m_D$ is found as follows~\cite{Kanemura:2014rpa, Han:2018zcn}:
\begin{align}
m_D=\frac{y_{\eta_{a}}V_{R_{ai}} M_i V^\dag_{L_{ib}} y_{S_{ib}} s_\alpha c_\alpha}{(4\pi)^2}
\left[\frac{m_{H_1}^2}{M^2_i-m_{H_1}^2}\ln\left[\frac{m_{H_1}^2}{M^2_i}\right]
-
\frac{m_{H_2}^2}{M^2_i-m_{H_2}^2}\ln\left[\frac{m_{H_2}^2}{M^2_i}\right]
\right],
\end{align}
The neutrino mass matrix is then diagonalized by a unitary matrix $U_\nu$ as \linebreak {$U_\nu^T m_\nu U_\nu ={\rm diag}(m_1,m_2,m_3)\equiv D_\nu$}.
Here we can identify $U_\nu$ as the PMNS matrix~\cite{Maki:1962mu} 
because of the diagonal mass matrix of the charged leptons, which is achieved by the $U(1)_{e-\mu}$ gauge symmetry.
Each of the mixings is then given by:
\begin{align}
\sin^2\theta_{13}=|U_{\nu_{13}}|^2,\quad 
\sin^2\theta_{23}=\frac{|U_{\nu_{23}}|^2}{1-|U_{\nu_{13}}|^2},\quad 
\sin^2\theta_{12}=\frac{|U_{\nu_{12}}|^2}{1-|U_{\nu_{13}}|^2}.
\end{align}

{
In case of NH, we find that
the neutrino mass eigenvalues and the effective neutrinoless double beta decay $\langle m_{ee}\rangle$ are respectively given in terms of observables and phases as
\begin{align}
&{m}_1^2  = 0,\quad {m}_2^2  = \Delta m^2_{\rm sol},\quad 
 {m}_3^2 \simeq  \Delta m^2_{\rm atm} 
 ,\\
& \langle m_{ee}\rangle \simeq
\left|\Delta m_{\rm sol} \sin^2\theta_{12} \cos^2\theta_{13}e^{i\alpha_{21}}
+ \Delta m^2_{\rm atm}  \sin^2\theta_{13}e^{i(-2\delta_{CP})}\right| ,
 \label{eq:NH-nu}
\end{align}
where $\Delta m^2_{\rm atm}$ and $\Delta m^2_{\rm sol}$ are respectively atmospheric mass difference square and solar mass difference square which are observables~\cite{Esteban:2018azc}; therefore these three neutrino mass eigenvalues are uniquely determined. 
Here, we redefine the neutrino mass eigenstate as {$|D_\nu|^2\equiv\frac{s^4_\alpha c^4_\alpha}{(4\pi)^8}{\rm diag}(0,|\tilde m_2|^2,|\tilde m_3|^2)$}. Then, $s_\alpha(c_\alpha)$ can be rewritten by
\begin{align}
s^4_\alpha c^4_\alpha =(4\pi)^8 \frac{\Delta m^2_{\rm atm}}{|\tilde m_3|^2},
\end{align}
which implies that $s_\alpha$ is determined by the two parameters $\Delta m^2_{\rm atm}$ and $|\tilde m_3|^2$.
Also, $\Delta m^2_{\rm atm}$ is fixed by 
\begin{align}
\Delta m^2_{\rm sol} =  \frac{|\tilde m_2|^2}{|\tilde m_3|^2}  \Delta m^2_{\rm atm}.
\end{align}


Similar to the case of NH,  we also find 
the neutrino mass eigenvalues and $\langle m_{ee}\rangle$ in case of IH to be
\begin{align}
&{m}_3^2  = 0,\quad {m}_2^2  = \Delta m^2_{\rm atm},\quad 
 {m}_1^2 =  \Delta m^2_{\rm atm}-\Delta m^2_{\rm sol},\\
 & \langle m_{ee}\rangle =
\left|\sqrt{\Delta m^2_{\rm atm}-\Delta m^2_{\rm sol}} \cos^2\theta_{12} \cos^2\theta_{13}
+\Delta m_{\rm atm} \sin^2\theta_{12} \cos^2\theta_{13}e^{i\alpha_{21}}\right|.
 \label{eq:IH-nu}
\end{align}
And $s_\alpha$ and $\Delta m^2_{\rm sol}$ are found by
\begin{align}
s^4_\alpha c^4_\alpha =(4\pi)^8 \frac{\Delta m^2_{\rm atm}}{|\tilde m_2|^2},\quad
\Delta m^2_{\rm sol} = \left(1- \frac{|\tilde m_1|^2}{|\tilde m_2|^2}\right)\Delta m^2_{\rm atm}.
\end{align}
Here, we redefine the neutrino mass eigenstate as $|D_\nu|^2\equiv\frac{s^4_\alpha c^4_\alpha}{(4\pi)^8}{\rm diag}(|\tilde m_1|^2,|\tilde m_2|^2,0)$.

}

\subsection{Lepton Flavor Violations} 
Lepton Flavor Violations (LFVs) arise from the term $y_\eta$ at one-loop level, and
their branching ratios are given by
\begin{align}
& {\rm BR}(\ell_a\to\ell_b\gamma)= \frac{48\pi^3\alpha_{\rm em} C_{ab} }{(4\pi)^4{\rm G_F^2}}
\left|\sum_{i=1-3}Y_{bi} Y^\dag_{ia} F(M_i,m_{\eta^-})\right|^2,\\
& F_{2}(m_a,m_b)=\frac{2 m_a^6 +3 m_a^4 m_b^2 -6 m_a^2 m_b^4 + m_b^6+12 m_a^4 m_b^2 \ln(m_b/m_a)}{12(m_a^2-m_b^2)^4},
\end{align}
where $Y \equiv  y_\eta V_R$, ${\rm G_F}\approx 1.17\times10^{-5}$[GeV]$^{-2}$ is the Fermi constant, $\alpha_{\rm em}\approx1/128$ is the fine structure constant at the Z-boson scale, $C_{21}\approx1$, $C_{31}\approx 0.1784$, and $C_{32}\approx 0.1736$.
Experimental upper bounds are respectively given by Refs.~\cite{TheMEG:2016wtm, Aubert:2009ag,Renga:2018fpd} as
\beq
{\rm BR}(\mu\to e\gamma)\lesssim 4.2\times10^{-13},\quad {\rm BR}(\tau\to e\gamma)\lesssim 3.3\times10^{-8},\quad {\rm BR}(\tau\to \mu\gamma)\lesssim 4.4\times10^{-8}
\label{LFVbounds}
\eeq
and these bounds give constraints on the related Yukawa couplings and masses in the loop.
It is worthwhile to mention the muon anomalous magnetic moment $\Delta a_\mu$.
Although we have a new contribution to $\Delta a_\mu$ from the same term as LFVs, its sign is negative, which is opposite to the experimental result. Thus, we assume a different effect to dominantly cause the anomaly and do not discuss it further.

\subsection{$Z_{e-\mu}$ Gauge Boson}  
After the $U(1)_{e-\mu}$ symmetry breaking, we find the massive $Z_{e-\mu}$ gauge boson that is denoted by $Z'$ hereafter, and its mass $m_{Z'}$ is given by
\begin{align}
m_{Z'}=g' v_\varphi,
\end{align} 
where $g'$ is the gauge coupling of the $U(1)_{e-\mu}$ symmetry and we neglect kinetic mixing for simplicity.
Gauge interactions among $Z'$ are given by
\begin{align}
&T =g' Z'_\mu(\bar e\gamma^\mu e - \bar\mu \gamma^\mu\mu+ \bar \nu_e \gamma^\mu P_L\nu_e- \bar \nu_\mu \gamma^\mu P_L\nu_\mu ) \label{eq:int-dm}
\\
&+ g' Z'_\mu
\sum_{i,j=1-3}\left(\frac12 \bar \psi_i (W_{N_L}+W_{N_R})_{ij} \gamma^\mu \psi_j
+\frac12\bar \psi_i (W_{N_L}-W_{N_R})_{ij} \gamma^\mu\gamma_5 \psi_j 
+ \bar \Psi_i W_{R_{ij}} \gamma^\mu P_R\Psi_j \right),\nn
 \end{align}
 where $W_{N_{L(R)}}\equiv V_{L(R)}^\dag {\rm Diag}[1,-1,0] V_{L(R)}$, and  $W_R\equiv U^\dag {\rm Diag}[1,-1,0] U$.
Then each of the decay rates of $Z'$ is given by 
 \begin{align}
& \Gamma(Z'\to e\bar e)\approx   \Gamma(Z'\to \mu\bar \mu)\approx  \Gamma(Z'\to \nu_{e,\mu}\bar \nu_{e,\mu})  \approx \frac{g'^2 }{12\pi} m_{Z'}, \\
& \Gamma(Z'\to X\bar X)\approx \frac{|(W_{N_L}+W_{N_R})_{11}|^2 g'^2 }{12\pi} m_{Z'} \left(1+\frac{m_X^2}{m_{Z'}^2}\right)\sqrt{\left(1-\frac{4m_X^2}{m_{Z'}^2}\right)},
\label{eq:dr1}
 \end{align}
 where we have 
assumed $2 M_1< m_{Z'} <M_{R_{1,2}}, M_2,M_3$ and $M_1$ is considered to be the DM in the next section. 
When the decay rate of $ \Gamma(Z'\to X\bar X)$ can be negligible, the branching ratios are respectively found as
 \begin{align}
& {\rm BR}(Z'\to e\bar e)\approx   {\rm BR}(Z'\to \mu\bar \mu)\approx  {\rm BR}(Z'\to \nu_{e,\mu}\bar \nu_{e,\mu})  \approx \frac13.
\label{eq:dr2}
 \end{align}

 Since $Z'$ couples to an electron and positron pair, we have to impose the following constraint which comes from LEP~\cite{Schael:2013ita}:
 \begin{align}
 4950\ {\rm GeV} \lesssim \frac{m_{Z'}}{g'},\label{eq:lep}
 \end{align}
 where we have adopted a conservative bound.
Here, we briefly mention other possibilities to detect signatures at colliders in the future, for the case of the $Z'$-mass being of the order of 100~GeV.
First is the Large Hadron Collider (LHC), which can observe the mode {$q\bar q\to Z/\gamma\to e^{+}e^-Z'(\mu^+\mu^-Z') \to \{2e^{+}2e^-,2\mu^{+}2\mu^-,e^{+}\mu^-e^{-}\mu^+\}$~\cite{Sirunyan:2018nnz}}. Second is the future International Linear Collider (ILC) which could measure modes {$e^+ e^-\to Z'\to \{e^+ e^-,\mu^+\mu^-\}$}~\cite{Baer:2013cma, Fujii:2017vwa}.
So far there is no analysis of LHC data for above channels, thus LHC provides no constraint on the model parameters.

\section{Properties of Dark Matter Particles} 
\label{DMprop}
The DM candidate in this model is the lightest Dirac fermion $X\equiv \psi_1$, and its mass given by $m_X\equiv M_1$.
In this section we study with which model parameters DM consisting of X and $\mathrm{\bar X}$ is viable, with the goal of showing the existence of an allowed region,
leaving a complete scan of the whole possible parameter space for future work.
First, we briefly discuss detectability by direct detection searches and the reason why we take $s_a=0$.
Then, we 
explain the calculation of the DM relic density which is determined by gauge interaction via s-channel, and perform a numerical analysis to 
explore the region around the pole $m_X=m_{Z'}/2$ which satisfies all discussed constraints.
Finally, we discuss that by applying Breit-Wigner enhancement to our model, the annihilation cross section in the current Universe can be increased by a boost factor $B$ 
compared to a generic thermally produced DM with velocity independent annihilation cross section.

\subsection{Direct Detection}
The latest bound on spin-independent scattering is reported by the XENON1T experiment, which gives an upper limit on the spin independent elastic DM-nucleon cross section \linebreak $\sigma$:  $\sigma<4.1\times 10^{-47}$ cm$^2$ at $m_X=30$ GeV with 90\% confidence level~\cite{Aprile:2018dbl}.
Our DM dominantly interacts with nuclei only via the mixing of $s_a$ at tree level arising from the terms $f_\varphi^{(')}$.  Then, our scattering cross section is given by
\begin{align}
\sigma\approx \frac{\mu^2_{nX}}{2\pi v^2} m_n^2 C^2|(V^\dag_R f'_\varphi V_L)_{11}|^2 (c_a s_a)^2\left(-\frac{1}{m_{h_1}^2} + \frac{1}{m_{h_2}^2} \right)^2,
\end{align}
where $\mu_{nX}\equiv m_n m_X/(m_n+m_X)$, $m_n$ is the mass of neutron, and $C\approx 0.3$ is determined by lattice simulation among DM and nucleon. 
The easiest way to evade this constraint is to assume $s_a$ to be zero.
LHC results also favor $s_a$ to be small with an upper bound of $s_a\lesssim 0.2$~\cite{Heinemeyer:2013tqa}. 
Thus, with the choice of $s_a=0$, no direct detection bounds need to be considered for our DM candidate. We leave exploring possible bounds for the case $s_a>0$ for future work.

\subsection{Relic Density}
In the following we discuss the relic density of DM.
The relevant processes arise from Yukawa interaction via $y_\eta$ and kinetic interaction via $g'$.
In case of Yukawa interaction, the coupling is mainly restricted by $\mu\to e\gamma$, which is typically of the order 0.01, although it depends on its flavor structure. Then the cross section via $y_\eta$ is $10^{-17}$ GeV$^{-2}$ at most.
Thus, Yukawa contribution is negligibly small compared to the cross section $\sim10^{-9}$~GeV$^{-2}$ required to explain the relic density.
As a result, the dominant cross section to the relic density comes from kinetic interaction. 

We make use of the micrOMEGAs package~\cite{Belanger:2013oya} to calculate the speed averaged cross section $\langle\sigma v_{\rm rel}\rangle$, and the relic density.
micrOMEGAs is adapted to this model by defining the properties of the interaction mediated by $Z'$ in the form of a kinetic term simplified from Eq.~\eqref{eq:int-dm} as follows:
\begin{align}
&T =g' Z'_\mu(\bar e\gamma^\mu e - \bar\mu \gamma^\mu\mu+ \bar \nu_e \gamma^\mu P_L\nu_e- \bar \nu_\mu \gamma^\mu P_L\nu_\mu + a_x \bar \psi_1 \gamma^\mu \psi_1)  \label{eq:int-dm-mm}
\end{align} 
where 
\begin{align}
 a_x = \frac12 |(W_{N_L}+W_{N_R})_{11}| .
\end{align} 
The model parameter space is thus effectively given by $m_X$ , $m_{\rm Z'}$ , $g'$  and  $a_x$ , with  $a_x$ taking values in the interval~$[0,1]$.

The evolution of the DM abundance is given through the Boltzmann equation
\begin{align}
\frac{dY}{dx}=-
\frac{x s[x] }{H}\langle\sigma v_{\rm rel}\rangle (Y^2-Y_{\rm EQ}^2), \label{eq:boltz}
\end{align}
where $s[x]$ is the entropy density and $H$ is the Hubble parameter, which are respectively given by
\begin{align}
s[x]=\frac{2\pi^2 g_\star}{45}\frac{M_X^3}{x^3},\quad
H= \sqrt{\frac{4\pi^3 g_\star}{45}} \frac{M_X^2}{M_{\rm PL}}.
\end{align}
Here $g_\star\approx$ 107 is the total number of effective relativistic degrees of freedom,
and the Planck mass \linebreak{$M_{\rm PL}\approx 1.22\times 10^{19}$~GeV}.
Finally, the DM relic density is given by
\begin{align}
\Omega h^2\approx 2.74\times 10^8  \left[ \frac{M_X}{\rm GeV}\right] Y_\infty,
\end{align}
where $Y_\infty$ is the final DM abundance~\cite{Srednicki:1988ce, Edsjo:1997bg, Guo:2009aj, Ibe:2008ye}. 
%
Observed relic density at $2 \sigma$ is given by Ref.~\cite{Ade:2013zuv} as
\begin{align}
\Omega h^2=0.1199\pm0.0054.
\end{align}
In the numerical analysis, we adopt a rather relaxed value range, $0.11\lesssim \Omega h^2\lesssim 0.13$, and the LEP constraint expressed in Eq.~\eqref{eq:lep} is imposed. 

The Breit-Wigner effect causes a higher DM annihilation rate than for a thermally produced DM with velocity independent annihilation cross section where average velocity $v$ is low, notably in the galactic DM halo near the position of the Solar system ($v \sim 10^{-3}$), and in the era of CMB formation ($v \sim 10^{-6}$).
To express this enhancement, we define the boost factor as the ratio of the speed averaged cross section of our model at a given value of $v$ under the condition of obtaining the correct relic density by solving Eq.~\eqref{eq:boltz}, and the speed averaged cross section required to obtain the correct relic density for a thermally produced DM with velocity independent annihilation cross section, given by  
\begin{align}
\langle\sigma v_{\rm rel}\rangle_{\rm th}\approx 3\times 10^{-26} {\rm cm^{3} s^{-1}}= 2.573\ 10^{-9}\ {\rm GeV}^{-2}.
\end{align}
The boost factor for annihilation in the galactic halo near the Solar System, in the current epoch is then given by
\begin{align}
B = \frac{\langle\sigma v_{\rm rel}\rangle_{\rm c}}{\langle\sigma v_{\rm rel}\rangle_{\rm th}},
\end{align}
and the boost factor for annihilation in the CMB formation era by
\begin{align}
B_{\rm CMB} = \frac{\langle\sigma v_{\rm rel}\rangle_{\rm CMB}}{\langle\sigma v_{\rm rel}\rangle_{\rm th}}, 
\end{align}
where $\langle\sigma v_{\rm rel}\rangle_{\rm c}$ corresponds to {$x = v^{-2} \approx 10^6$}, while $\langle\sigma v_{\rm rel}\rangle_{\rm CMB}$ corresponds to $x\approx 10^{12}$~\cite{footnote2}.

%
\subsection{Numerical Analysis}
We have performed a numerical analysis to find the allowed region for obtaining the correct relic density of DM, where neutrino oscillation data is implicitly reproduced and the LFV constraint is satisfied. 
We have analyzed parameter sets with fixed values of  {$m_X=[20,100,400,2000]$ GeV}, while $m_{Z'}$ and $g'$ are determined by randomly selected values of $\delta\equiv 1-\frac{m_{Z'}^2}{4m_X^2}$ and $\gamma\equiv \frac{\Gamma_{Z'}}{m_{Z'}}$, with the mixing matrix $M$ also being chosen randomly under the condition of $m_X$ being the smallest eigenvalue, from which we calculate the effective input parameter $a_x$. To cover this parameter space, O$(10^7)$ parameter sets are calculated per value of $m_X$ with a flat distribution in the range {[-8,-3]} of both  $\log_{10}(-\delta)$ and $\log_{10}(\gamma)$. 
The properties of the parameter sets satisfying $0.11\lesssim \Omega h^2\lesssim 0.13$ and $4950\ {\rm GeV} \lesssim \frac{m_{Z'}}{g'}$ are further analyzed. 

We also conduct a numerical analysis on LFVs for these parameter sets, finding that if the Yukawa couplings are below $\sim 10^{-2}$, 
the experimental limits given in Eq.~\eqref{LFVbounds} are not exceeded for any of the parameter sets. 
Given that the Yukawa couplings are independent from the parameters defining the DM properties, there is no constraint from LFV on the studied parameter space.  

\begin{figure*}
\begin{center}
\includegraphics[width=0.49\linewidth]{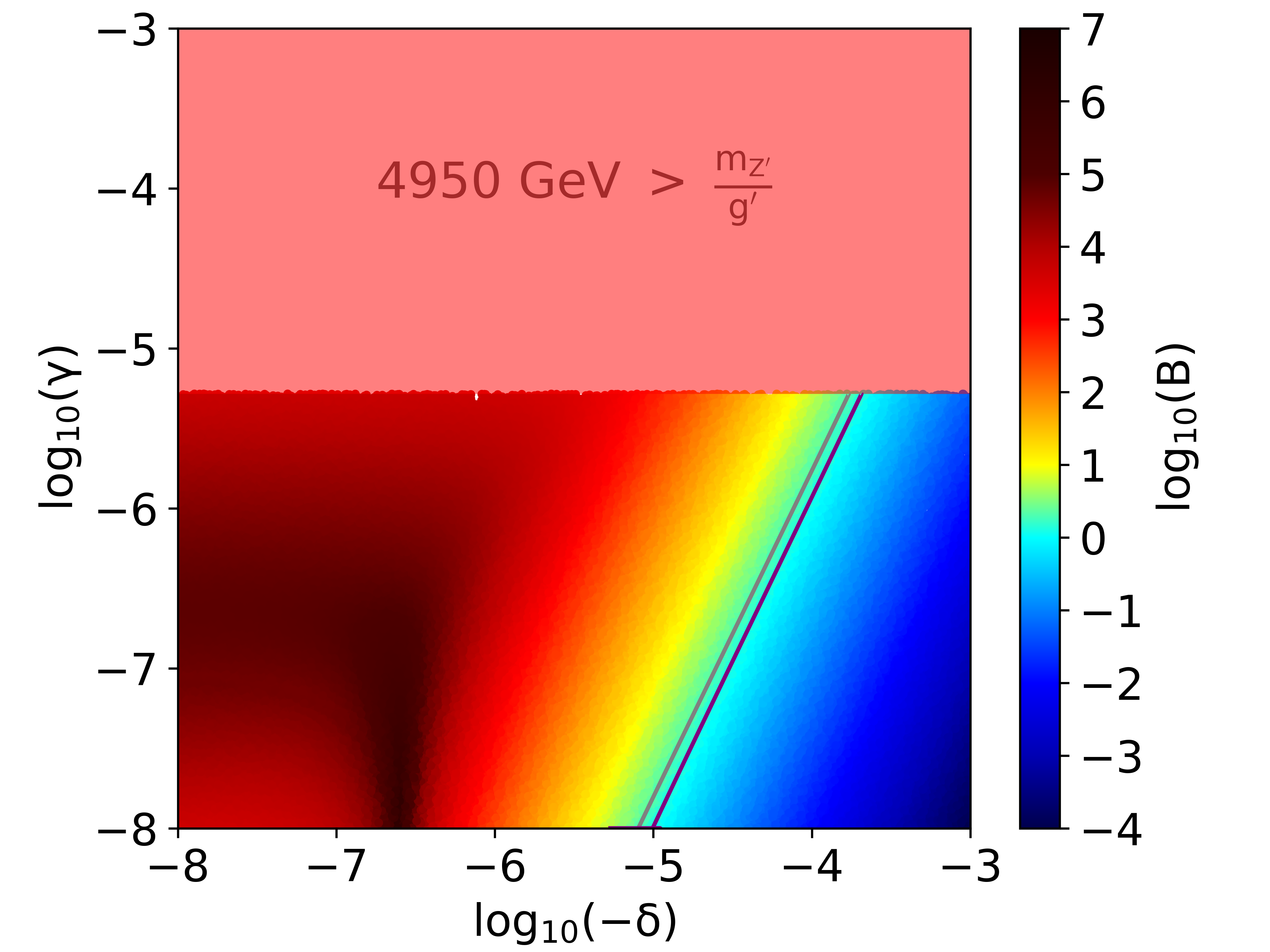} 
\includegraphics[width=0.49\linewidth]{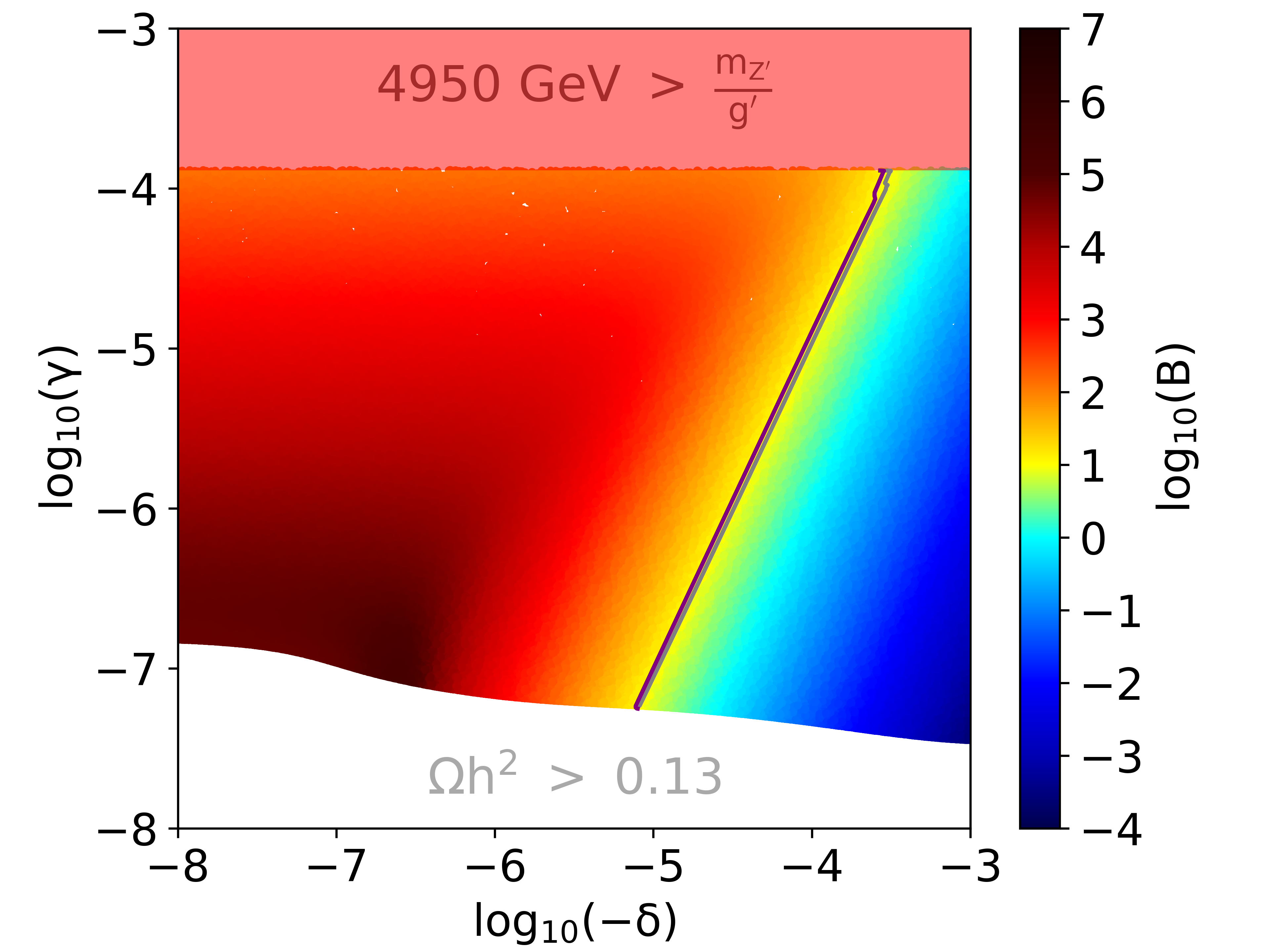}
\includegraphics[width=0.49\linewidth]{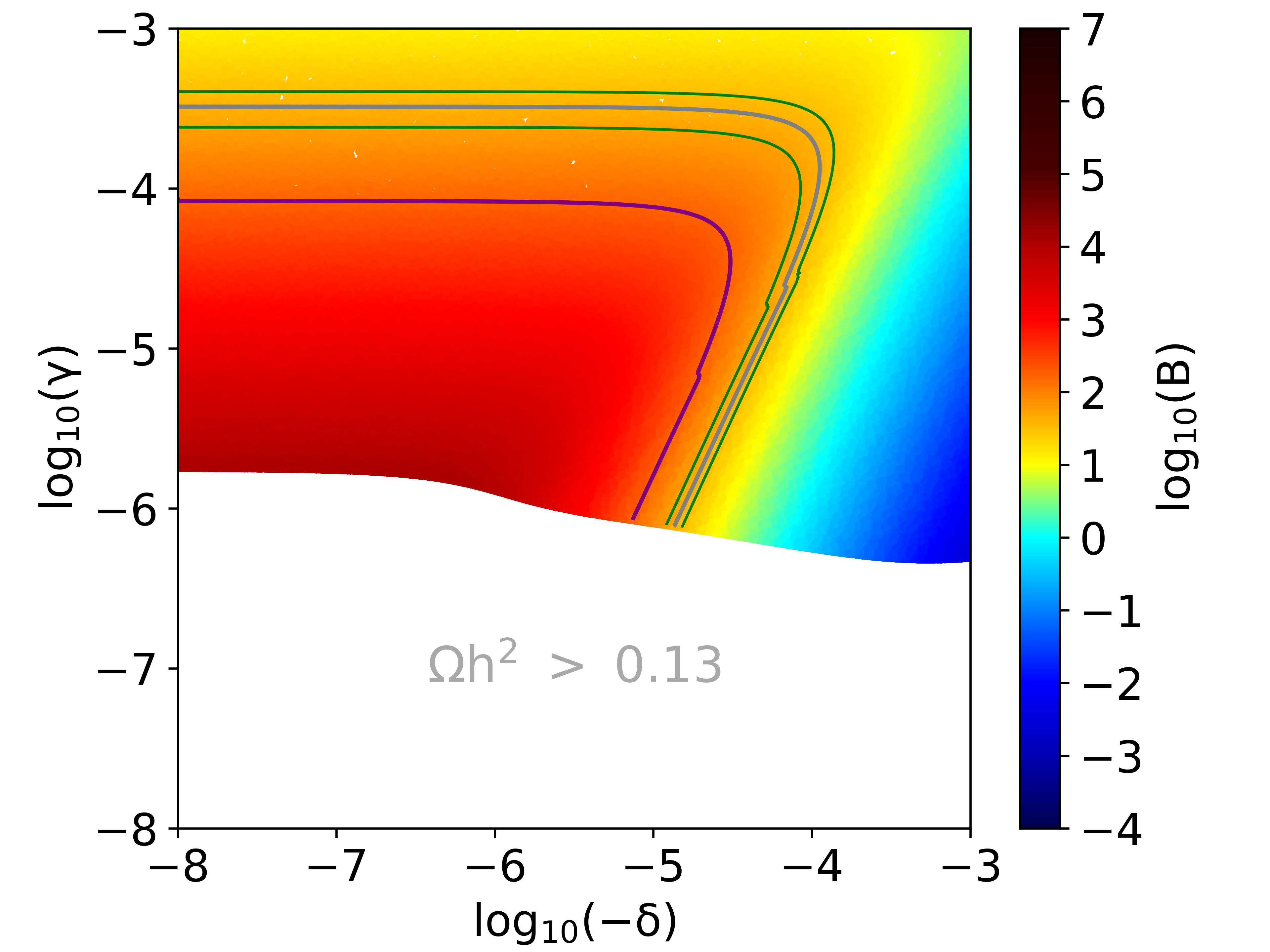} 
\includegraphics[width=0.49\linewidth]{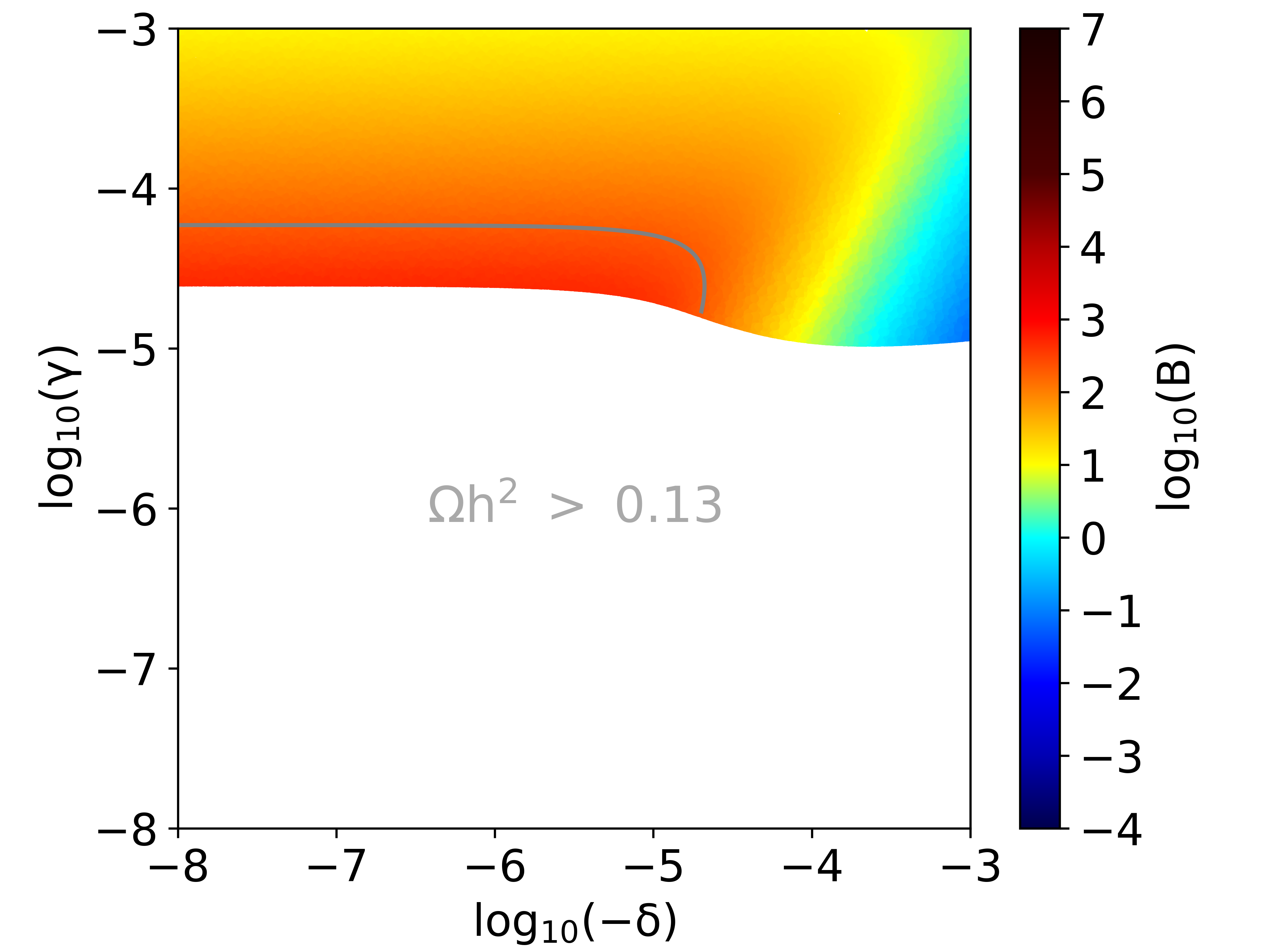}

\caption{Scatter plots in terms of $-\delta$ and $\gamma$ of the parameter sets fulfilling \mbox{$0.11 < \Omega h^2 < 0.13$} and \mbox{$4950\ {\rm GeV} \lesssim \frac{m_{Z'}}{g'}$}, where the left-top figure is the one for $M_X=20$~GeV, the right-top one for $M_X=100$~GeV, the left-bottom one for $M_X=400$~GeV, and the right-bottom one for~$M_X=2000$ GeV. Boost factor $B$ is encoded by the color of the dots.
The gray line indicates the limit from CMB anisotropy based on $B_{\rm CMB}$ using Eq.~\eqref{CMBconstr}, the purple lines the limit on $B$ from analysis of CALET and AMS-02 data in \secref{DMsig}, and the green lines in the plot for $m_X = 400$~GeV are the boundaries of the $2 \sigma$ region given in Figure~\ref{limits} for explaining a structure in the CALET spectrum as a signature of this DM candidate.
The region excluded by the LEP bound $ 4950\ {\rm GeV} \lesssim \frac{m_{Z'}}{g'}$ is colored red, while in the white region, $\Omega h^2 > 0.13$ for all parameter sets.
}
  \label{g-d}
\end{center}\end{figure*}

With micrOMEGAs, we calculate the speed averaged cross section for {$v = 10^{-3}$} and {$v = 10^{-6}$} to obtain $B$ and $B_{\rm CMB}$ for each parameter set respectively, taking  {$\langle\sigma v_{\rm rel}\rangle_{\rm th} = \frac{0.12}{\Omega h^2}\times  3\times 10^{-26} {\rm cm^{3} s^{-1}}$} to compare with the generic model yielding the same relic density.
Figure~\ref{g-d} shows boost factor $B$ for the parameter sets which satisfy $0.11\lesssim \Omega h^2\lesssim 0.13$ and $ 4950\ {\rm GeV} \lesssim \frac{m_{Z'}}{g'}$ in the $\log_{10}(-\delta)$ vs. $\log_{10}(\gamma)$ plane, indicating that these two parameters determine the value of $B$. The left-top plot is for $m_X=20$ GeV, the right-top plot for $m_X=100$ GeV, the left-bottom plot for $m_X=400$ GeV, and the right-bottom plot for $m_X=2000$ GeV. 

Among the cosmological constraints to the model, CMB anisotropy provides the strictest bound, since in principle the annihilation rate increases with decreasing relative velocity~\cite{Hisano:2011dc}. The limit calculated from 2015 Planck CMB anisotropy measurement~\cite{Kawasaki:2015peu} excludes 
\begin{align}
\langle\sigma v_{rel}\rangle_{th} > \frac{m_X}{GeV} \times 1.4 \times 10^{-27} {\rm cm}^3 {\rm s}^{-1}
\end{align}
 for annihilation to \EP, and 
\begin{align}
\langle\sigma v_{rel}\rangle_{th} > \frac{m_X}{GeV} \times 3.6 \times 10^{-27}{\rm cm}^3 {\rm s}^{-1}
\end{align}
 for annihilation to $\mu^{-} + \mu^{+}$. While velocity dependence of the annihilation cross section was not considered for these results, it can be assumed that the annihilation cross section in the CMB formation era is decisive. With the limits in principle being inversely proportional to the energy injected into the thermal bath, the limit for the annihilation of $X$ can be calculated as 
\begin{align}
B_{\rm CMB} > (\frac{1}{3}\times 1.4^{-1} + \frac{1}{3}\times 3.6^{-1})^{-1} \times \frac{10^{-27}}{3 \times 10^{-26 }} \times \frac{m_{\rm X}}{\rm GeV} . \label{CMBconstr}
\end{align}
For example, at $m_X = 400$~GeV, $B_{\rm CMB} > 40.32$ would be excluded, with the part of the parameter space excluded by this and corresponding limits for other values of $m_X$ indicated in Figure~\ref{g-d}.

\begin{figure*}
\begin{center}
\includegraphics[width=0.49\linewidth]{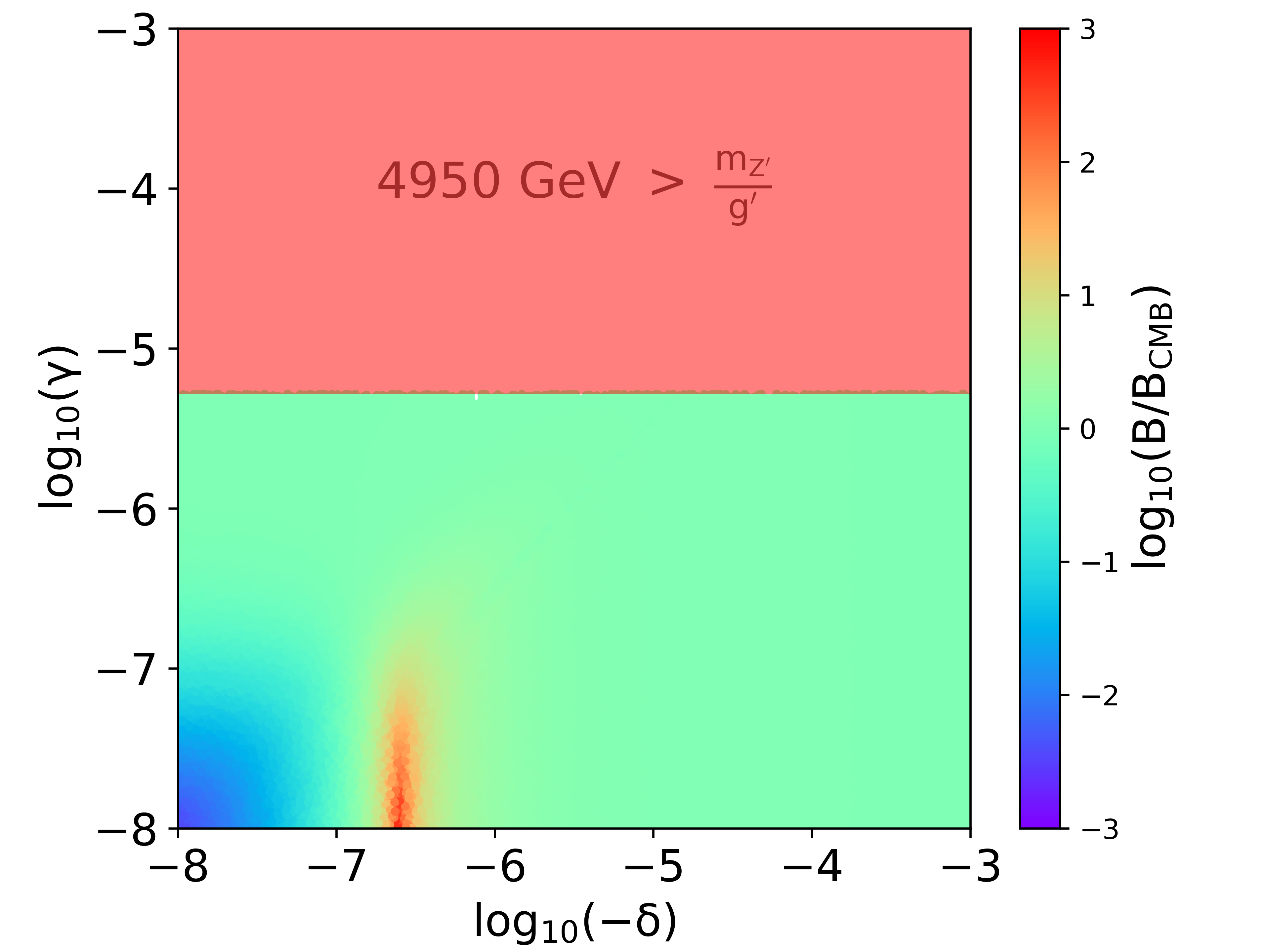} 
\includegraphics[width=0.49\linewidth]{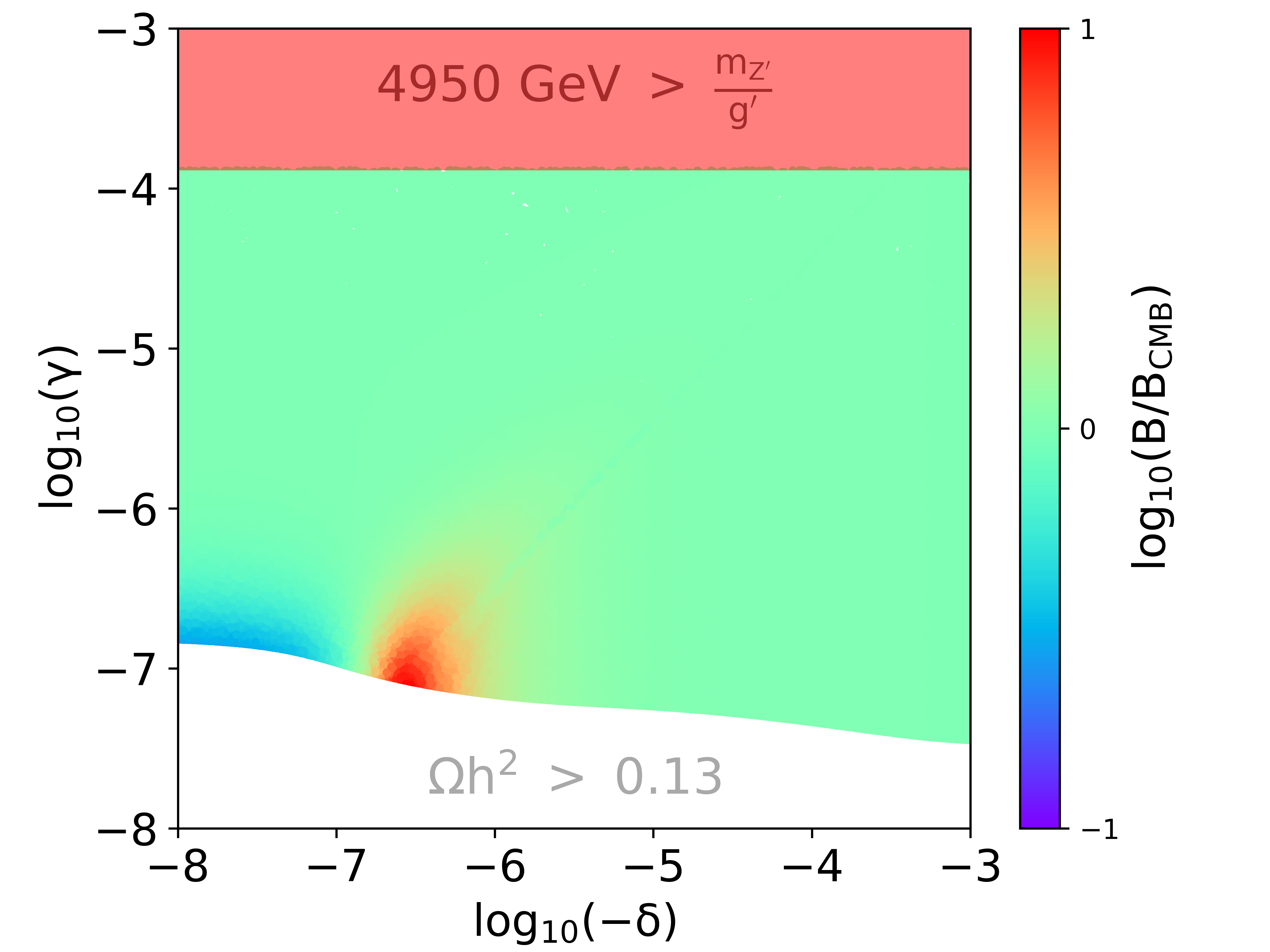} 

\caption{Scatter plots showing the ratio of the boost factor in the galactic halo $B$ to the boost factor at the time of CMB formation $B_{\rm CMB}$  encoded by the color of the dots. The left figure is the one for $m_X=20$~GeV, the right one for $m_X=100$~GeV. We omit the figures for $m_X=400$~GeV and $m_X=2000$~GeV, since there is no allowed region where $B \not\approx B_{\rm CMB}$ for these values of $m_X$.
} 
  \label{br}

\end{center}\end{figure*}

We find $B_{\rm CMB} \approx B$ for most of the studied parameter space, except for two regions at $\gamma  \lesssim 10^{-7}$ as shown in Figure~\ref{br}, matching the results shown in Ref.~\cite{Xiang:2017jou}. While there is a region in which $B$ is up to three orders of magnitude larger than $B_{\rm CMB}$, it is ruled out by the constraint on $B_{\rm CMB}$.

\section{Dark Matter Signatures in Cosmic-Ray Spectra}
\label{DMsig}
Electron and positron cosmic rays from annihilation of the DM candidate particle X are a potential signature of the proposed model.  
In this section we evaluate measured cosmic-ray spectra by CALET and AMS-02 to derive limits on the annihilation cross section and to identify potential correlations of spectral structures with the DM signature.  The results of CALET and AMS-02 agree well for the \EP~spectrum, which is a prerequisite for the combined fitting of the CALET \EP~spectrum and the AMS-02 $e^{+}$-only spectrum without assuming an inherent systematic offset. Due to the systematic difference of the DAMPE \EP~spectrum results~\cite{Ambrosi:2017all} from both AMS-02 and CALET spectra, we chose not to consider them in our study.

\subsection{Electron and Positron Flux from Annihilation in the Galactic Halo}

To predict the shape of the spectral component from annihilation of X and $\mathrm{\bar X}$ , the positron spectra (identical to electron spectrum due to the symmetry of the process) per annihilation in the electron and muon channels have been calculated with PYTHIA 8.2~\cite{Sjostrand:2014zea}, which in turn were used as input for the propagation calculation with {DRAGON~\cite{Gaggero:2013rya}} to obtain the flux at Earth. For the local DM density, $\rho_0 = 0.3 $~GeV/cm$^3$ is assumed, and the speed averaged annihilation cross section normalized to the value predicted for a thermal relic DM,  $\langle\sigma v_{rel}\rangle_{th} = 3 \times 10^{-26}$~cm$^3$~s$^{-1}$. The choice of the DM halo shape model has no strong impact on the spectrum as the propagation range of electrons is limited and discussed models agree around the position of the solar system~\cite{Weber:2009pt}. A NFW parametrization~\cite{Navarro:1996gj} is used for the calculation. The flux for annihilation of X and $\mathrm{\bar X}$ is composed according to the branching ratio from Eq.~\eqref{eq:dr2} as the sum of one third of the normalized flux for electron channel and one third of the flux for muon channel, with the annihilation to neutrinos not contributing.

\pagebreak
For the propagation calculation, we consider two propagation models strongly distinct in diffusion zone height $L$ and diffusion coefficient normalization $D_0$, denoted Model~A and Model~B. 

Model A comprises a gradual change in the slope of the diffusion coefficient with rigidity~\cite{Genolini:2017dfb} according to \beq D(R) = D_0 \left(\frac{R}{R_0}\right)^{\delta_l} \left(1 + \left(\frac{R}{R_b}\right)^{\frac{{\delta_l-\delta_h } }{s}} \right)^{-s} \ , \label{DReq}\eeq with $\delta_l = 0.62 $ , $\delta_h = 0.33 $, $ R_0 = 4 $~GV , $R_b = 350 $~GV , $D_0 = 1.1 \times 10^{28}\ \mathrm{cm^2/s}$, and a softness parameter $s = 0.15$. These propagation parameters are derived from calculation of the nuclei spectra with DRAGON. Setting the diffusion zone half-height $L = 3$~kpc and the width of the spiral arm thickness to 0.65~kpc, this model reproduces the AMS-02 B/C ratio~\cite{Aguilar:2016vqr} and proton spectrum~\cite{Aguilar:2015ooa} measurements if assuming a common injection index {$\gamma_{i}$ = -2.32} for all nuclei. This model predicts the hardening in the proton spectrum matching the index change as recently measured by CALET~\cite{Adriani:2019aft} as a pure propagation effect, without any break in the injection index.

Model B is designed as an alternative with high diffusion coefficient already at low energy, choosing $D_0 = 3.7 \times 10^{28}\ \mathrm{cm^2/s}$ which implies a much larger diffusion zone half-height of $L = 15$. With a constant diffusion coefficient index $\delta_h = 0.5$, the slope changes in B/C ratio and proton spectrum are explained as the effects of diffusive acceleration (Alfven speed $v_A = 12$~km/s) and two smooth breaks in the nuclei injection spectrum at 12~GV and 500~GV, changing the power law index from 2.0 to 2.36 and from 2.36 to 2.1 respectively. Here, the spiral arm width is taken as the default value of 0.3~kpc.

\begin{figure*}

\centering
	\resizebox{0.95\linewidth}{!}{\includegraphics{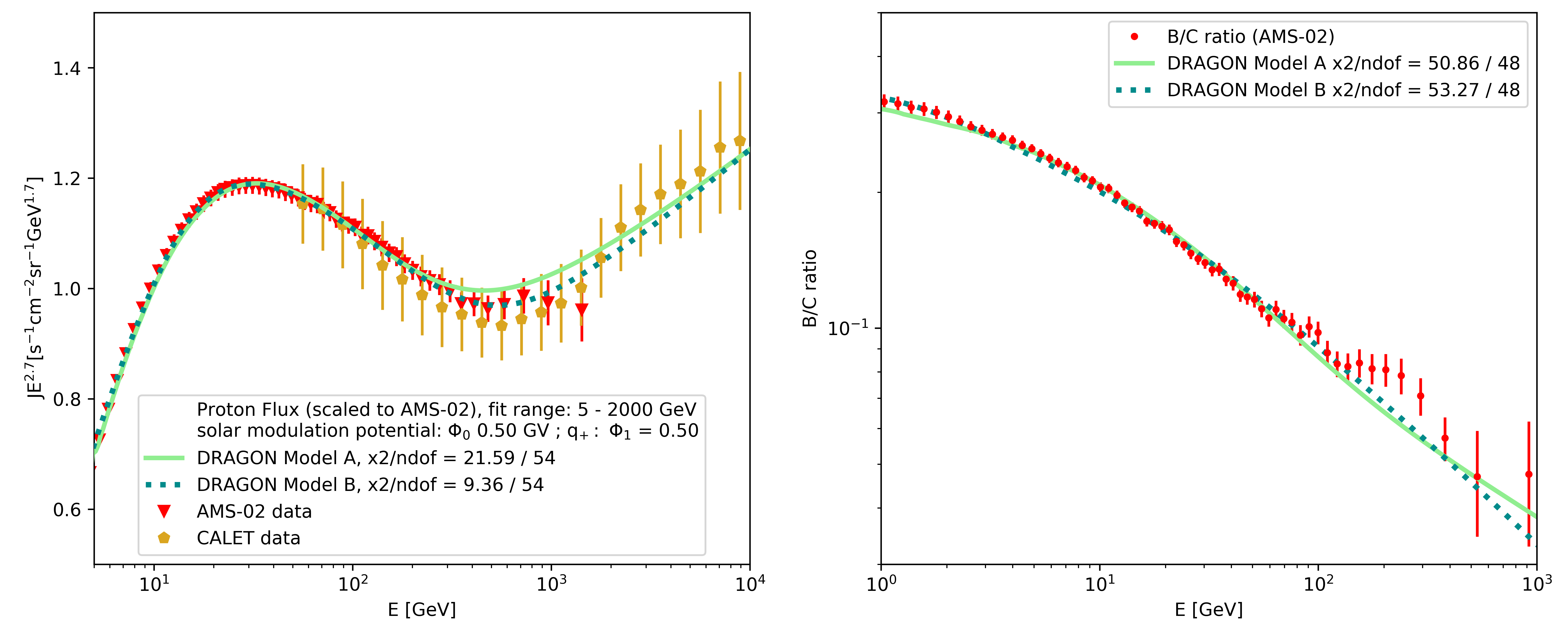}}  
\begin {center}
\caption{Proton flux and B/C ratio as reproduced by propagation Model A and Model B compared to data from AMS-02 and CALET, with charge independent solar modulation potential {$\Phi_0= \Phi = 500$~MV}.  $\Phi_1$ represents an additional potential for positive charge only at low energy following Ref.~\cite{Cholis:2015gna}. \label{propmods}} 
\end {center}
\end{figure*}

Figure~\ref{propmods} shows the comparison of the calculated nuclei spectra for both propagation models with experimental data.

\subsection{Astrophysical Background Flux Model}

Potential signatures of DM in the electron and positron spectra need to be distinguished from the background spectrum from astrophysical sources. The three main components comprising the background spectra are primary electrons accelerated by supernova remnants (SNR), secondary electrons and positrons from interaction of nuclei cosmic rays with the interstellar medium, and an extra source of electrons and positrons as an explanation of the positron excess, for which nearby pulsars are the prime candidate~\cite{Hooper:2008kg,Malyshev:2009tw,Kawanaka:2009dk,Feng:2015uta}. The pulsar scenario is supported by the discovery of {$\gamma$-ray} emission around nearby pulsars~\cite{Hooper:2017gtd,Fang:2018qco} and thus chosen over other discussed explanations of the positron excess such as secondary production in dense clouds around SNRs~\cite{Blasi:2009hv,Fujita:2009wk,Blasi:2009bd,Mertsch:2009ph,Blum:2013zsa,Cholis:2013lwa,Kohri:2015mga}. While a DM-only explanation of the positron excess is also not ruled out, it requires specific conditions such as decaying dark matter yielding softer spectra than the electron and muon channel annihilation of our DM candidate~\cite{Buch:2016jjp,Bhattacharyya:2017kfq,Farzan:2019qdm}.

The model used for describing the background spectra and fitted to electron and positron cosmic-ray data is the sum of the above mentioned components, with the electron spectrum written as 
\beq \Phi_e^{-} = C_e E^{-(\gamma_e - \Delta\gamma_e)} \left(1+ \left( \frac{E}{E_b}\right)^{\frac{\Delta\gamma_e}{s}} \right)^s e^{-\left(\frac{E}{E_{cut_d}}\right)} +\frac{C_s}{C_{norm}} \Phi_{s(e^{-})} +  \Phi_{ex}  \ , \eeq 
and the positron spectrum as 
\beq \Phi_e^{+} = \frac{C_s}{C_{norm}} \Phi_{s(e^{+})} +  \Phi_{ex} \ . \eeq

Due to their large energy loss in propagation, the spectrum of primary electrons depends on the distribution of individual SNR in the galactic neighborhood of the solar system, which is yet mostly unknown. As an effective model of the local (after propagation) primary electron spectrum from all contributing SNR, it is parametrized by a power law with a soft spectral break (normalization $C_e$, index $\gamma_e$, break position $E_b$ and index change $\Delta\gamma_e$ are free fit parameters, softness $s = 0.05$ is fixed) at low energy, and a high-energy exponential cut-off at $E_{cut_d}$ representing radiative energy loss of high energy electrons. $E_{cut_d}$ is not well constrained by the measurement and therefore treated as a fixed nuisance parameter for which we consider values of 2~TeV, 4 TeV and 10~TeV. 

The secondary positron ($\Phi_{s(e^{+})}$) and electron ($\Phi_{s(e^{-})}$) fluxes are taken from the output of numerical propagation calculation with DRAGON for the nuclei spectra from which the propagation conditions are derived. The propagation conditions are used consistently for calculation of fluxes from secondaries, pulsars and Dark Matter. With an initial scale factor $C_{norm}$ obtained from normalizing the proton flux to measurements of AMS-02~\cite{Aguilar:2015ooa}, a common rescaling factor ($C_s/C_{norm}$) is included in the fit as free parameter to account for remaining uncertainties in secondary particle production. 

For the flux of the primary positron source causing the positron excess $\Phi_{ex}$, the least complex solution of a single young pulsar is assumed in the base model, for which the Monogem pulsar (PSR J0659+1414) is chosen.
The power-law with cut-off injection spectrum (defined by spectral index $\gamma_{ex}$ and cut-off energy $E_{cut_{ex}}$) of the pulsar is propagated using the analytic solution of the propagation equation for a point source as explained e.g. in Ref.~\cite{Feng:2015uta}, adapted to include the gradual change in the diffusion coefficient of propagation Model A, yielding the propagated flux from the pulsar as  
\beq \Phi_{pulsar} = \frac{Q_0 \eta}{\pi^{3/2} r^3_{dif}} E^{-\gamma_{ex}} \left(1-\frac{E}{E_{max}} \right)^{(\gamma_{ex}-2)} e^{-\frac{E/E_{cut_{ex}}}{1-E/E_{max}}-\frac{r^2}{r^2_{dif}}} \ ,  \eeq
in which the characteristic diffusion distance $r_{dif}$ is expressed as 
\beq r_{dif} = 2 \sqrt{\frac{D(E) t_{dif}}{1-\delta(E)} \frac{E_{max}}{E} \left[1-\left(1-\frac{E}{E_{max}} \right)^{(1-\delta(E))} \right]}\ , \eeq
with $E_{max}= (b_0\ t_{dif})^{-1}$, $b_0=1.4\ 10^{-16}$~GeV~s$^{-1}$, $D(E)$ given by Eq.~\eqref{DReq} and $\delta(E)$ approximated as the local index of $D(E)$ at $E$. 

The distance to the Monogem pulsar $r = 0.28$~kpc is taken from the ATNF catalog~\cite{Manchester:2004bp}, as well as its age $T = 1.11 \times 10^2$~kyr and energy loss rate $\dot E = 3.81 \times 10^{34}$~erg~s$^{-1}$. The initial rotation energy of the pulsar $Q_0 = 1.48 \times 10^{48}$~erg is calculated as $Q_0 = \dot E\ T^2 / \tau$, where $\tau = 10$~kyr is the assumed spin-down timescale~\cite{Feng:2015uta}, so that the spectrum scales with the acceleration efficiency $\eta$, which is a free parameter in the fitting. A common assumption is that the accelerated particles are trapped for some time in the pulsar wind nebula (PWN) forming around the pulsar, and released with the dissolution of the PWN.  The release delay $T_r$ is thus subtracted from the age $T$ of the pulsar to determine the diffusion time $t_{dif}$, with $T_r$ scanned in steps of 1~kyr considering the range up to 100~kyr~\cite{Malyshev:2009tw}. 

\pagebreak
\subsection{Fit of the Background Model to CALET and AMS-02 Data}
The model is fitted to the data of CALET~\cite{Adriani:2018ktz} based on total flux $\Phi_e^{-}+\Phi_e^{+}$ and data of \mbox{AMS-02}~\cite{Aguilar:2019owu} for $E>10$~GeV based on $\Phi_e^{+}$ by minimizing the sum of \X2 of both comparisons, with systematic uncertainties of both measurements taken into account. For the CALET measurement, the $1 \sigma$ deviation $\Delta_{(k,i)}$ as a function of each data point's energy ($E_i$) is listed in the supplemental material of Ref.~\cite{Adriani:2018ktz} for the systematic uncertainty associated with the following parts of the analysis:  Normalization, tracking, charge selection, electron identification, Monte Carlo model dependence. 
A systematic shift of the data-points is performed as part of the fit function with weights $w_k$ as free parameters and the squared weight of each uncertainty is added to the total \X2 of the fit as given by
\beq \chi^2_{CALET} = \left( \sum\limits_{i} \frac{(\Phi_i) + \sum_k \Delta_{(k,i)} w_k - J_{i})}{\sigma_i^2} \right) + \sum\limits_k w^2_k \ , \eeq
where $i$ iterates over the data points and $k$ over the different systematic uncertainty types. Systematic errors associated with the trigger and the boosted decision tree proton rejection are added quadratically to the statistical error.
For the AMS-02 measurement, the error on mean energy $\sigma_E$ in each bin is translated into an error on flux $\sigma_{J(E)}$ using the power law index $\gamma_{e^+}$ also shown in Ref.~\cite{Aguilar:2019owu} via the relation $\sigma_{J(E)} = J(E) (\sigma_E/E) (\gamma_{e^+}-1)$.

The lower boundary of $E>10$~GeV for the data points used in the fitting is chosen due to charge and time dependent solar modulation effects expected below this energy~\cite{Cholis:2015gna}. Solar modulation effects above this energy are calculated using the force field approximation with a modulation potential of $\Phi = 500$~MV, common for both charge signs. To check the potential influence of the parameter on our results, $\Phi = 300$~MV and $\Phi = 700$~MV are used as alternative fixed values.

\begin{figure*}[!htb]
\centering
	\resizebox{\linewidth}{!}{\includegraphics{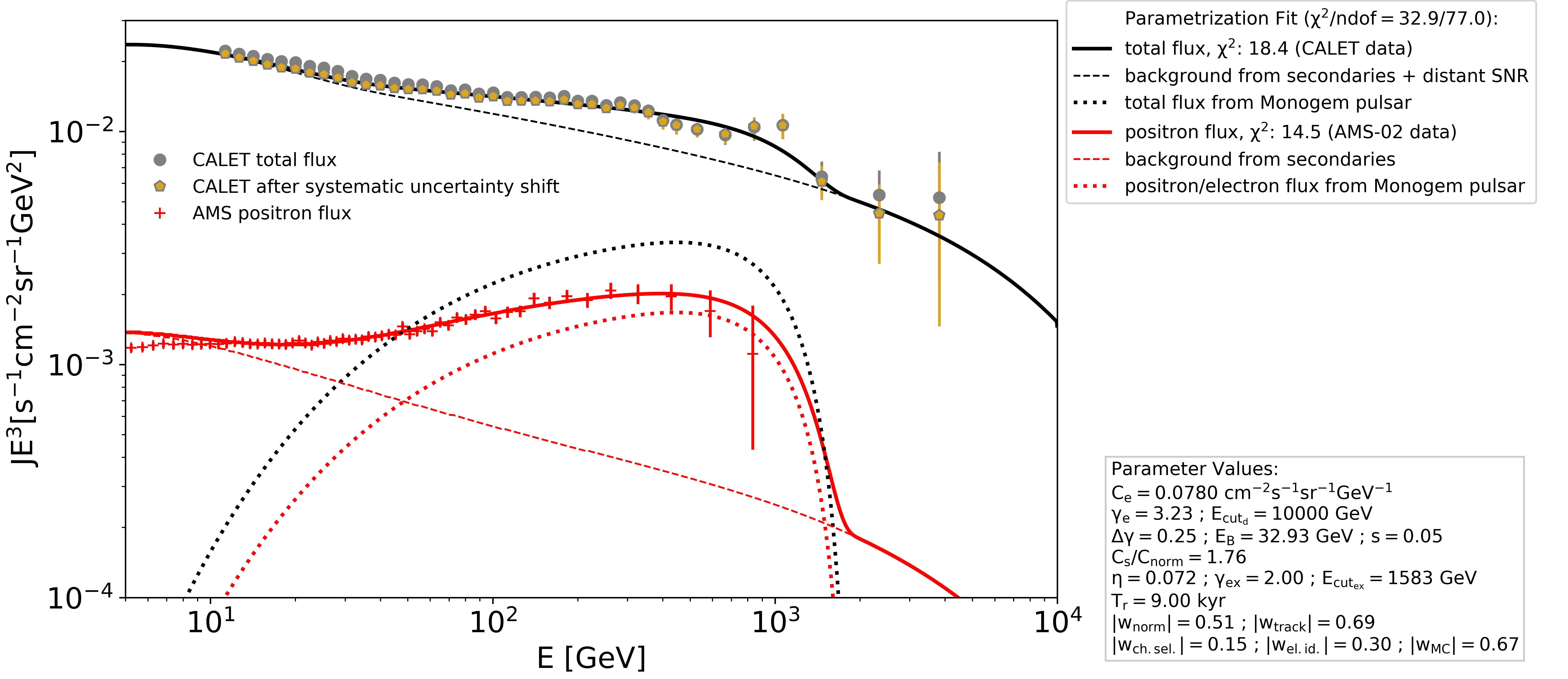}}  	    
	\resizebox{\linewidth}{!}{\includegraphics{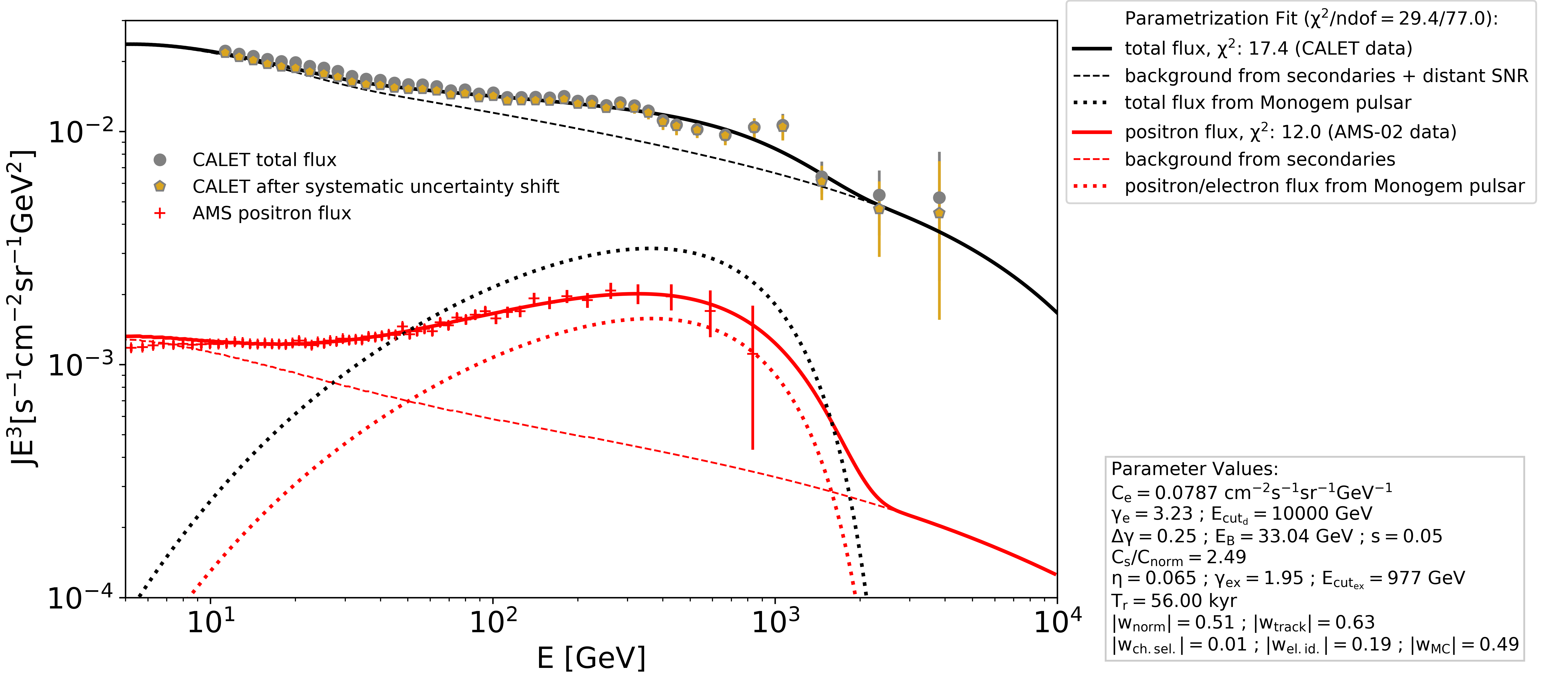}}  	    
\begin {center}
\caption{The base model as fitted to CALET and AMS-02 data using propagation Model A (top) and Model B (bottom) in the default case of $E_{cut_{d}} = 10$~TeV and $\Phi = 500$~MV. See legends for explanation of markers and lines, the values for the parameters introduced in the text are given in the box to the right of the graphs. \label{basemod}} 
\end {center}
\end{figure*}

The results of the base-model fit are shown in Figure~\ref{basemod} for the two propagation models. For Model A, the best fit is obtained with $T_r = 9 $~kyr, while for Model B, $T_r = 56$~kyr yields lowest \X2. The best fit for both propagation models uses $E_{cut_{d}} = 10$~TeV, which is thus taken as the default case. The reduced \X2 is in either case $\chi^2/ndof \approx 0.5$, indicating that the base model already more than adequately describes the data.  With the Geminga pulsar as source of the positron excess, the fit quality is significantly worse unless $T_r>100$~kyr, which is the reason why we chose the Monogem pulsar.

\FloatBarrier

\subsection{Limit on Annihilation cross section from CALET and AMS-02 Data} 

\begin{figure*}[!hptb]
\centering
	\resizebox{\linewidth}{!}{\includegraphics{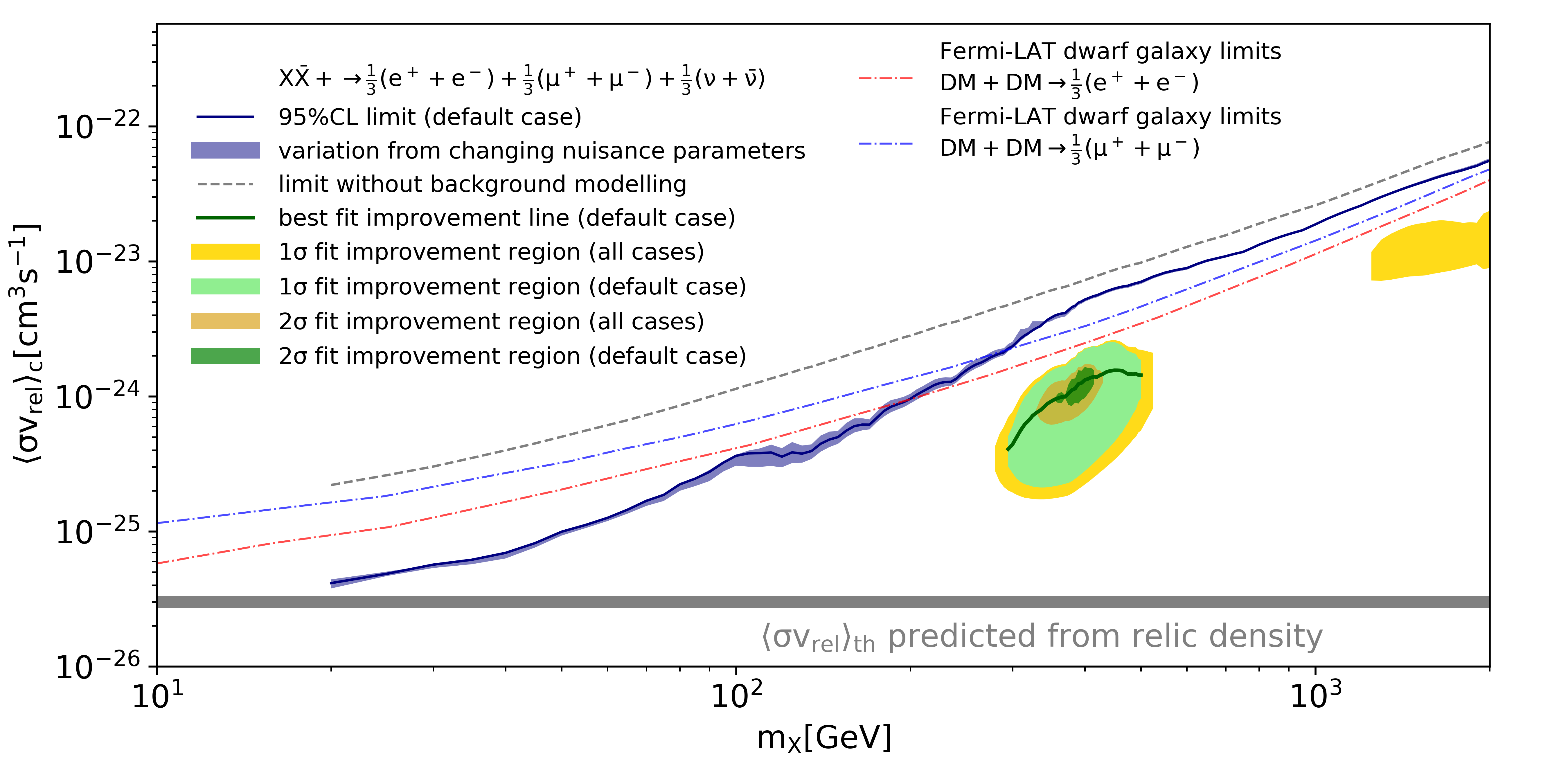}}  	    
	\resizebox{\linewidth}{!}{\includegraphics{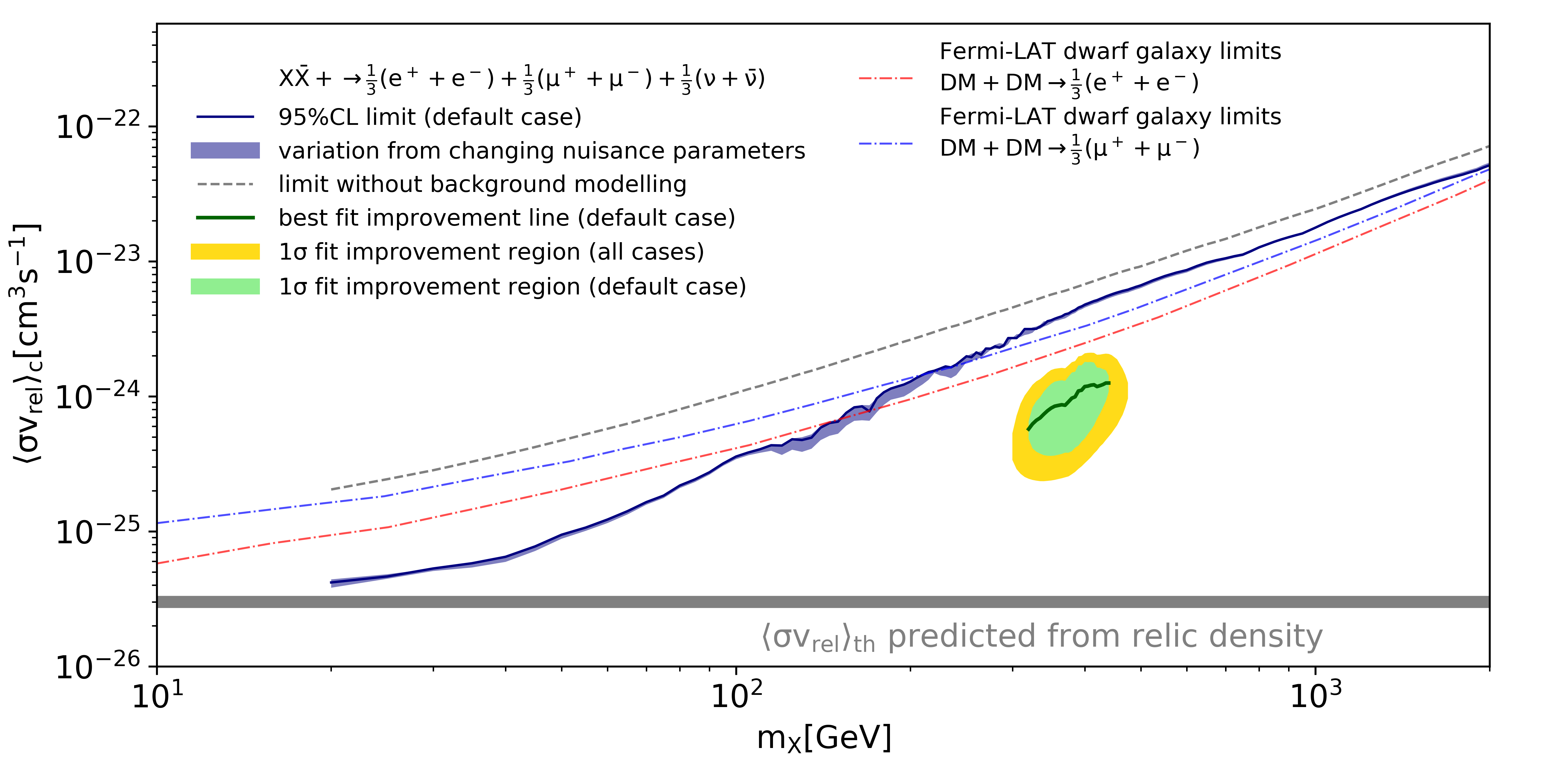}}
\begin {center}
\caption{Limit on the annihilation cross section as a function of $m_X$, compared to limits for $e^\pm$ and $\mu^\pm$ channels from {$\gamma$-ray} observation of dwarf galaxies with Fermi-LAT from the supplemental material of Ref.~\cite{Ackermann:2015zua} multiplied with three to account for the branching fraction. The purple area shows the variation of the limit among all cases with nuisance parameters $\Phi$ and $E_{cut_{d}}$ changed between 300~MV,500~MV,700~MV and 2~TeV,4~TeV,10~TeV respectively. The dashed gray line indicates the limit without any background modeling, with only excess over measured flux contributing to \X2. Also, the shown limit is the worst in the range from variation of $\Phi$. The dark green line shows the cross section as a function of $m_X$ of the best fit where $\Delta \chi^2 > $2.2977 ($1 \sigma$). The green and yellow areas show the areas with $1 \sigma$ fit improvement for the default case and all cases respectively. The dark yellow and dark green areas show the areas with $2 \sigma$ fit improvement. The top panel is for propagation Model A, the bottom panel for Model B. \label{limits} }
\end {center}
\end{figure*}

The predicted flux from DM annihilation is added to the base model as an additional component of $\Phi_{ex}$ with varied boost factor $B$, and the change of \X2 studied.
To derive a limit on $B$, or equivalently the speed averaged annihilation cross section $\langle\sigma v_{rel}\rangle_{c}$, $B$ is increased in steps until \X2 exceeds the 95\%~CL threshold for the respective number of degrees of freedom~\cite{footnote3}.
To determine the precise value of $B$ for which the 95\%~CL threshold is crossed, the scan is repeated from the last allowed value with a factor 10 smaller step size, down to a step size of 0.01. To avoid reporting a too stringent limit due to the fitting function having no unique minimum, the "Migrad" and "Simplex" minimizers of Minuit~\cite{James:1994vla} are used in alternation with different starting points as explained in Ref.~\cite{Motz:2015cua}. Multiplying the normalization cross section by the scale factor at which the 95\%~CL threshold is crossed yields the limit on cross section $\langle\sigma v_{rel}\rangle_{c}$. 
By performing this procedure with $m_X$ scanned in steps of 5~GeV up to 500~GeV, 25~GeV from 500 to 1~TeV, 50 GeV from 1~TeV to 2~TeV and 100~GeV above 2~TeV, limits depending on $m_X$ are calculated, which are shown in Figure~\ref{limits}. It is found that the limit varies only slightly under change of the nuisance parameters $\Phi$ and $E_{cut_{d}}$.
In principle, these limits are subject to the modeling of the astrophysical background flux being a good representation of the actual spectrum, disregarding whether it is an correct interpretation e.g. whether or not the Monogem pulsar is indeed the dominating source of the positron excess. However to judge the conservativeness of the limits, it should be considered that for the peaked DM signal to be hidden by structures of the background from multiple astrophysical sources, these structures would have to form a deficit in a rather narrow energy range which can be considered an implausible coincidence given the smoothness of the spectrum in general.   
Due to these reasons, we consider the limits rather conservative, however to estimate the utmost possible influence of the background variability, also limits without any assumption on the background were calculated, using the method described above, but with only excess of the flux from DM annihilation over the flux measured by CALET and AMS-02 contributing to \X2. They are also shown in Figure~\ref{limits} for comparison.

\subsection{Structures in the CALET Spectrum as Possible Dark Matter Signatures}

The addition of the predicted DM flux and increase of $B$ for limit calculation causes eventually an increase in \X2 of the fit. However, it is found that the addition of the DM flux with $B$ smaller than the limit value improves the fit compared to the base model with a pulsar extra source in two ranges of $m_X$, corresponding to step-like structures in the CALET spectrum. 
Given the excellent energy resolution combined with detailed energy calibration~\cite{Asaoka:2017clb} over the wide dynamic range~\cite{Asaoka:2018ope} of CALET, 
it is permissible to assume that the measured structures are features of the physical spectrum and not measurement artifacts, thus warranting an interpretation.
To quantify the significance of interpreting the spectral structures as a signature of the proposed DM candidate, the optimal value of $B$ and associated maximal \X2 reduction are determined depending on $m_X$. Using an approach similar to the limit calculation, $B$ is initially scanned in 20 steps between zero and the 95\%~CL limit value, and then the interval around the value with best \X2 scanned in nested intervals. 


\begin{figure*}[t]

\centering
	\resizebox{0.495\linewidth}{!}{\includegraphics{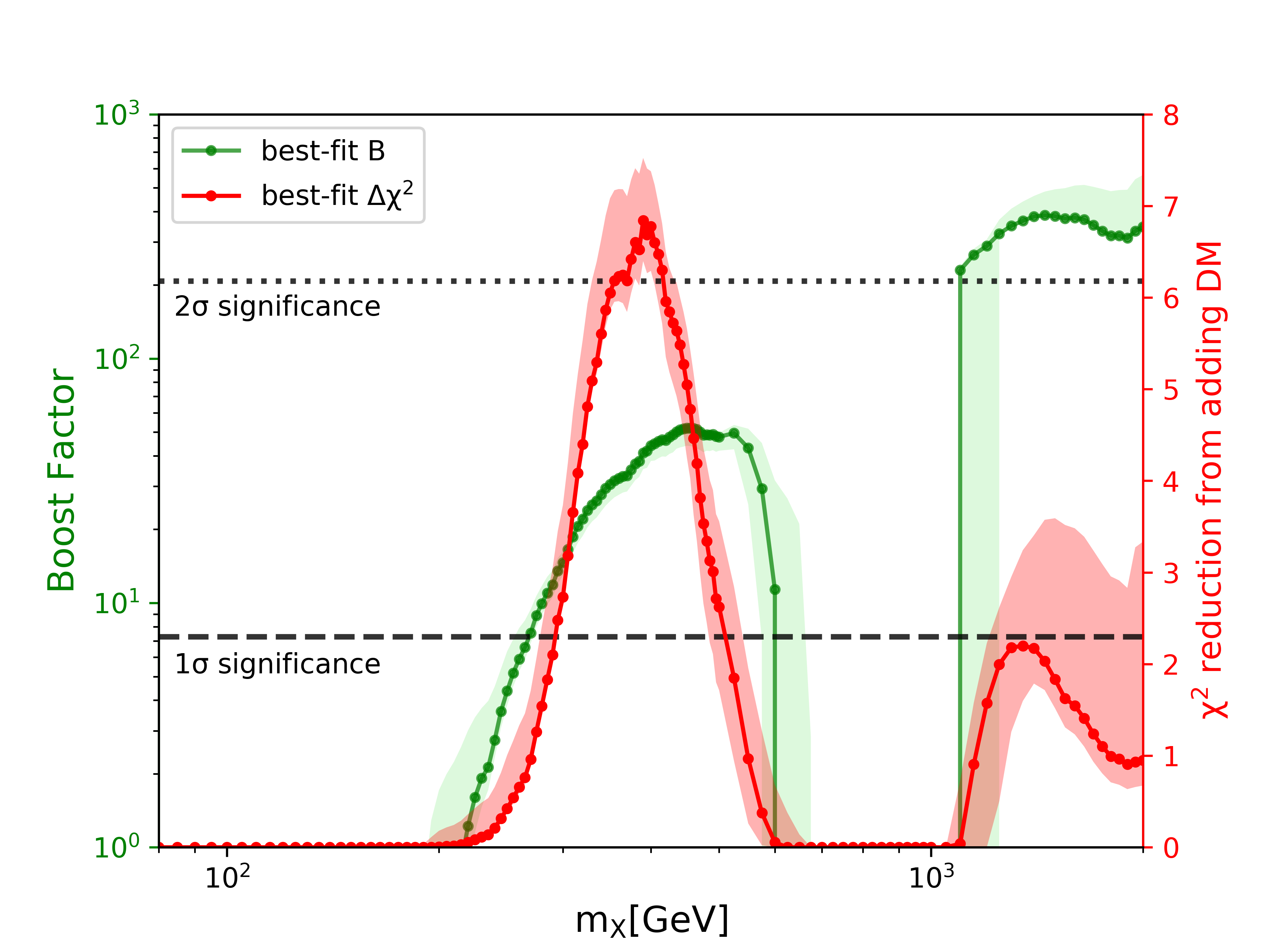}}  	    
        \resizebox{0.495\linewidth}{!}{\includegraphics{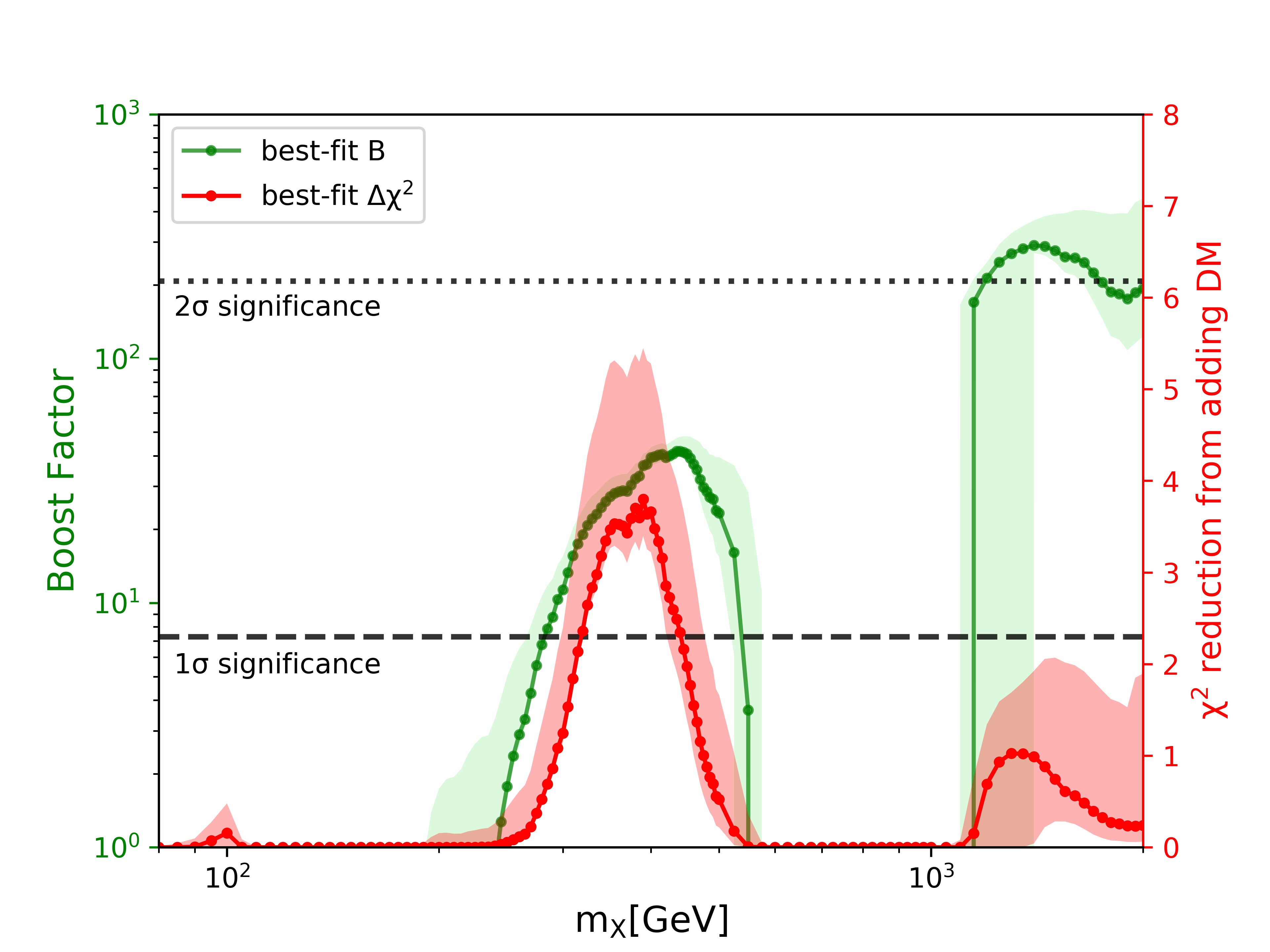}}  
	
\begin {center}
\caption{Fit improvement (\X2 reduction) by addition of flux from Dark Matter to the base model for propagation Model A (left) and Model B (right) as a function of $m_X$ (red), together with the values of $B$ giving the best fit (green). The shaded regions indicate the change from variation of the nuisance parameters.  \label{fitimp}} 

\end {center}
\end{figure*}


The best-fit $B$ and \X2 improvement as a function of DM mass are shown in Figure~\ref{fitimp}. The largest \X2 improvement compared to the single pulsar case is $\Delta\chi^2$ = 6.84 at $m_X = 390$~GeV with 
$B = 40.1$ 
for propagation Model A. The significance exceeds the $2\sigma$ significance level for the two additional free parameters ($m_X$ and $B$ or $\langle\sigma v_{rel}\rangle_{c}$) independent of the chosen values for the nuisance parameters. For propagation Model B, the best fit for the default case of $E_{cut_{d}} = 10$~TeV improves by $\Delta\chi^2$ = 3.80 also at $m_X = 390$~GeV, with larger improvement if choosing smaller $E_{cut_{d}}$. 
The necessary Breit-Wigner enhancement of $B \approx 40$ is predicted within the theoretical framework of the DM candidate as shown in Figure~\ref{g-d}, with part of the $2\sigma$-region being below the constraint from CMB anisotropy. It is also not ruled out by overproduction of {$\gamma$-rays} as shown by the comparison with limits from dwarf-galaxy observation by Fermi-LAT in Figure~\ref{limits}.

\begin{figure*}

\centering
	\resizebox{\linewidth}{!}{\includegraphics{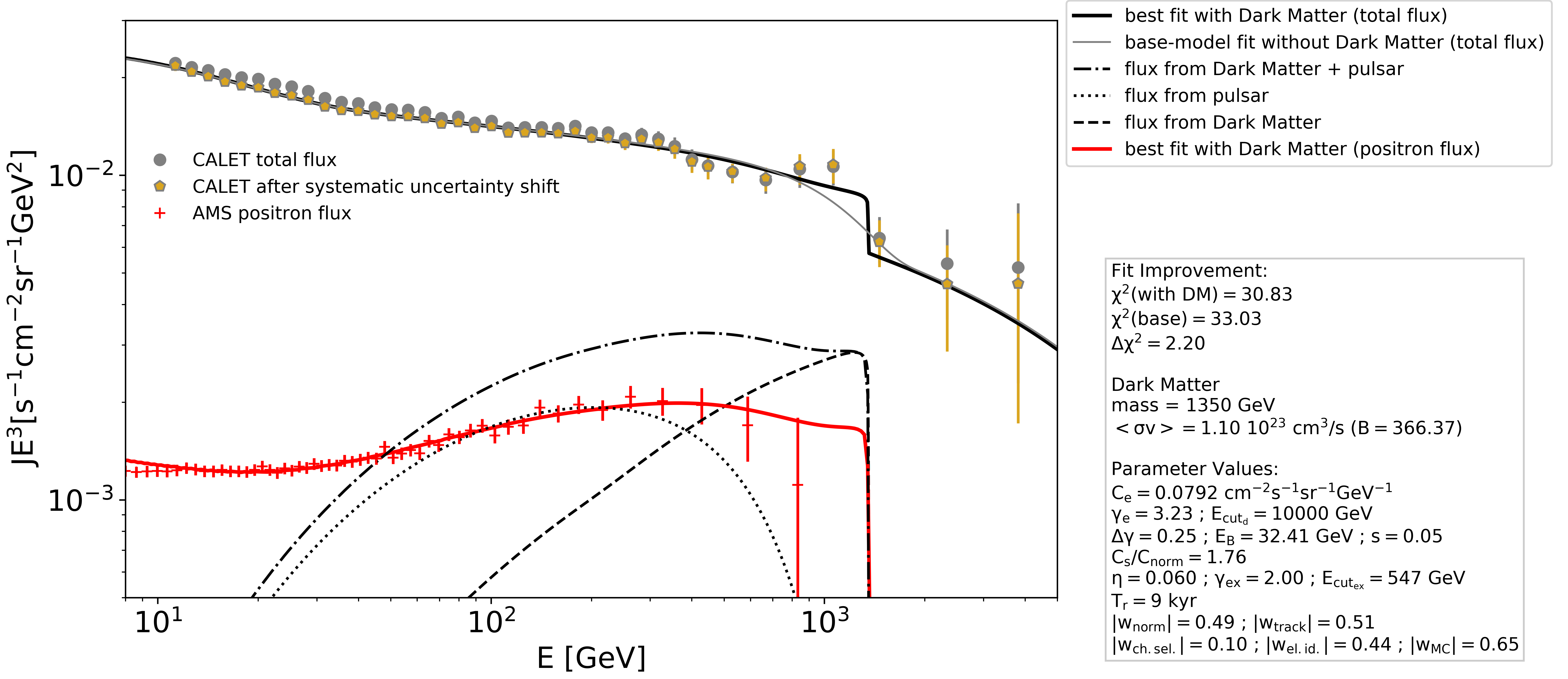}}
	\resizebox{\linewidth}{!}{\includegraphics{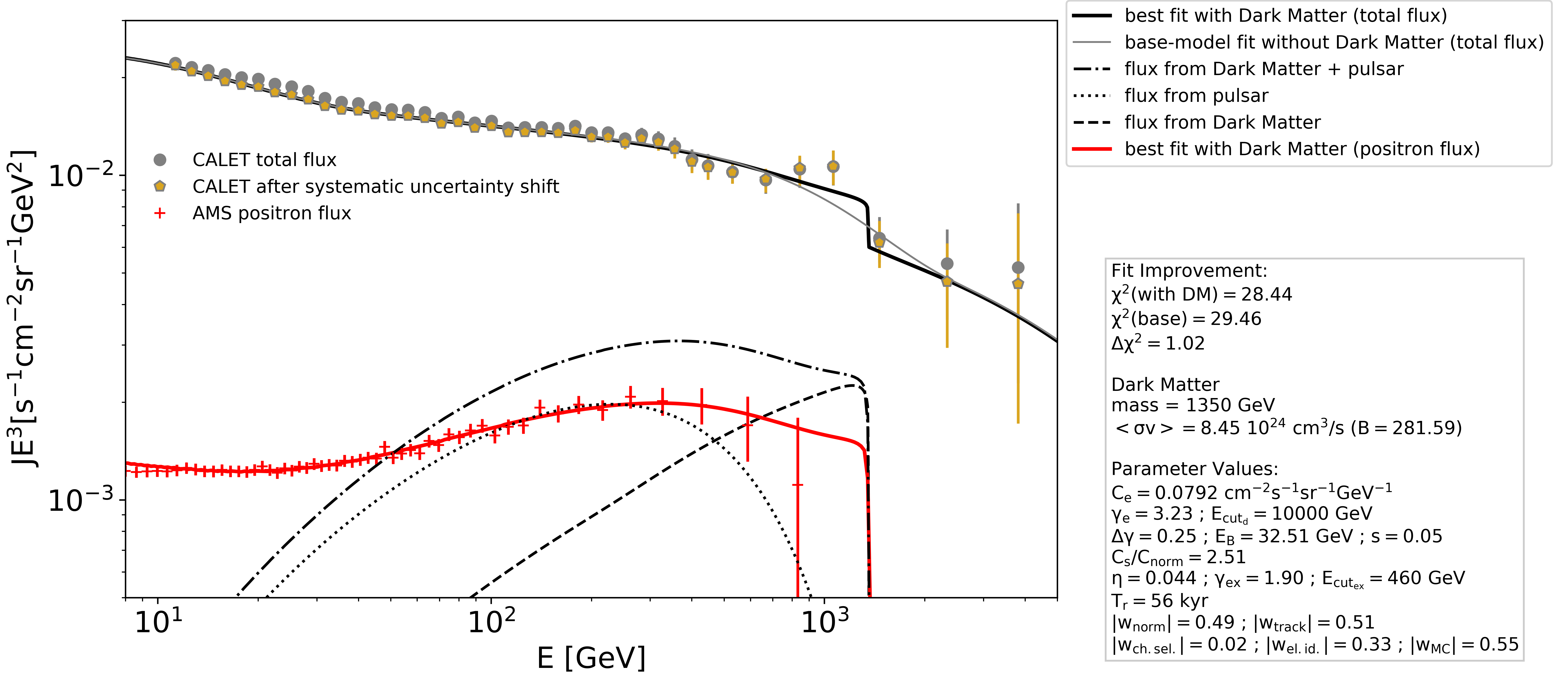}}  	  
\begin {center}
\caption{The best fits for the default case in the TeV region with $M_{\rm DM}= 1350$~GeV for propagation Model A (top) and Model B (bottom). See legends for explanation of markers and lines, the values for the parameters introduced in the text are given in the box to the right of the graphs. \label{bestfitsTeV}} 

\end {center}
\end{figure*}

With both propagation models, the fit also improves with addition of the DM signal above 1~TeV where another a step-like structure exists in the CALET spectrum. For the default case, the fit improvement is maximal at $m_X = 1350$~GeV for both Model A and Model B, but it is much less significant than the improvement for $m_X = 390$~GeV. Apart from the larger errors, the significance is low, since even for the direct annihilation to \EP, the DM signal spectrum is not localized (hard) enough to match the structure well, as the best-fit graphs in Figure~\ref{bestfitsTeV} demonstrate.
The structure at $m_X = 390$~GeV is thus a better candidate for being a DM signature, despite the 1~TeV structure being visually more prominent. 

\begin{figure*}

\centering
	\resizebox{\linewidth}{!}{\includegraphics{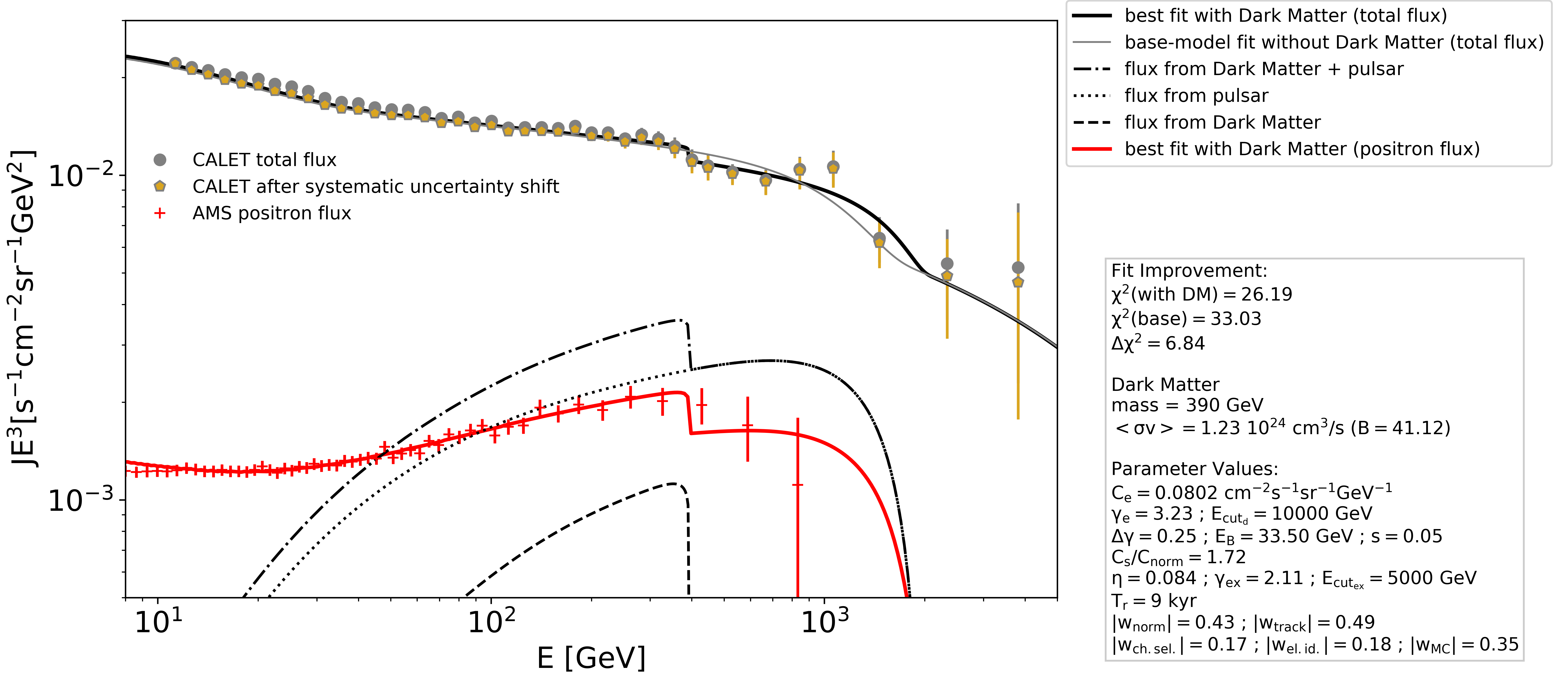}}
	\resizebox{\linewidth}{!}{\includegraphics{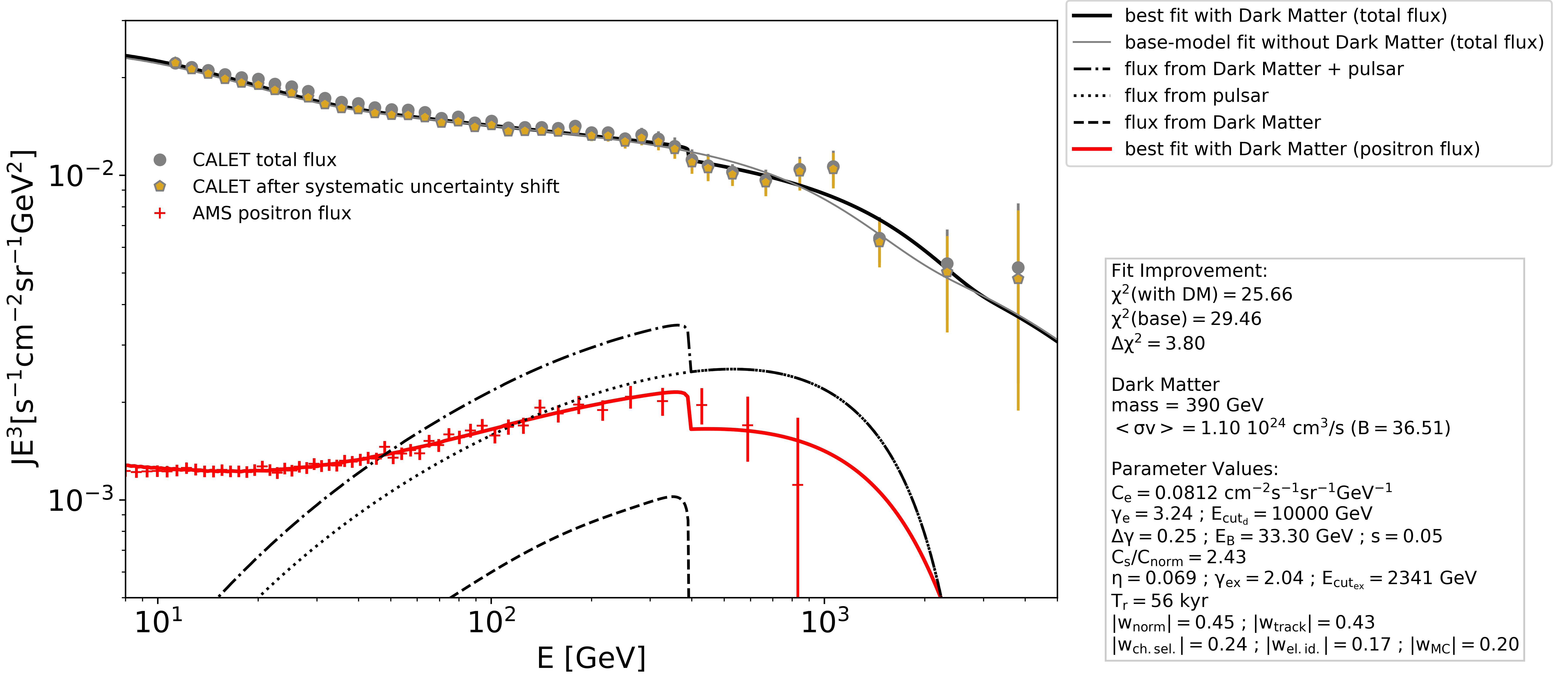}}  	  
\begin {center}
\caption{The best fits for the default case with $M_{\rm DM} = 390$~GeV for propagation Model A (top) and Model B (bottom). See legends for explanation of markers and lines, the values for the parameters introduced in the text are given in the box to the right of the graphs. \label{bestfits}} 
\end {center}
\end{figure*}

\clearpage
\section{Summary and Conclusions}
\label{sumconcl}

We have shown the viability of a GeV-TeV range WIMP-like DM candidate featuring flavor-dependent interaction only with electron and muon, allowed by current constraints from direct and indirect detection. The DM candidate is predicted in the framework of a scotogenic model at two-loop level, where we have accommodated two families of Dirac neutral fermions and Majorana fermions under gauge $U(1)_{e-\mu}\times Z_3\times Z_2$. The Dirac fermion with lightest mass is our DM candidate and it runs inside the neutrino loop, which is a typical feature of the scotogenic scenario.
The two families are the minimal extension to understand the neutrino oscillation data, its mass eigenstate and the gauge anomaly cancellations at the same time.
Due to introducing two families, we have predicted that the two mass eigenstates of active neutrinos are uniquely given by the two observed mass difference squares, $\Delta m^2_{sol}$ and $\Delta m^2_{atm}$ depending on the hierarchy order, with the lightest mass eigenvalue being always zero. 
Next, we have shown the allowed region yielding the correct relic density of DM in terms of $m_{X}$ and $m_{Z'}$, while imposing the constraint from LEP. Furthermore we have shown that the annihilation cross section may be increased by a boost factor $B$ from Breit-Wigner enhancement, depending on the two factors $\delta$ and $\gamma$, finding that $(|\delta|,\gamma)\lesssim {\cal O}(10^{-4}-10^{-3})$
can give ${\cal O}(10-100)\ B$.

After calculating the expected signature of the DM candidate in electron and positron cosmic rays for two largely distinct propagation models, we performed a combined search in the measured \EP~CALET and $e^+$ AMS-02 spectra on top of an astrophysical background model assuming a single young pulsar as the source of the positron excess.
As outcome we presented limits on the annihilation cross section close to those from {$\gamma$-ray} observation with Fermi-LAT, as well as a possible association of structures in the CALET spectrum with a DM signature. 
The significance of the fit improvement by adding the DM signature to the base model exceeds the $2\sigma$-level depending on the propagation model, with the best fit for $M_{\rm DM}$ at 390~GeV with a value of $B$ which is well within the range predicted by the Breit-Wigner enhancement.  These results demonstrate the significance of the step-like structure itself, and while other interpretations are possible, for example by overlapping spectra from individual astrophysical sources~\cite{Motz:ICRC2019}, it is shown that the annihilation of the DM candidate from the model presented herein also provides a suitable explanation. 

\pagebreak
\section*{Acknowledgments}
The research performed by H.M. leading to the results shown in this paper was supported by a Waseda Grant for Special Research Projects (Project number: 2020C-539).
The research performed by H.O. was supported by an appointment to the JRG Program at the APCTP through the Science and Technology Promotion Fund and Lottery Fund of the Korean Government. It was also supported by the Korean Local Governments - Gyeongsangbuk-do Province and Pohang City (H.O.). H. O. is sincerely grateful for being a KIAS member, and to Log Cabin at POSTECH for providing a nice space to
come up with ideas for this project. K.K. was supported by JSPS KAKENHI Grants No. JP17H01131, MEXT Grant-in-Aid for Scientific Research on Innovative Areas JP15H05889, JP18H04594, JP19H05114. This work was also supported in part by KAKENHI 19H04617 (Y.A.) 


\begin{thebibliography}{99}
\bibitem{footnote2}
Depending on the parameters, the velocity dependence may also cause a reduction of the annihilation rate for low velocity, i.e. $B < 1$.
\bibitem{footnote3}
Requiring the absolute threshold value for \X2 instead of an threshold for the increase of \X2 should be considered the conservative choice given that the reduced \X2 of the base model is 	significantly less than one.
\end{thebibliography}
\end{document}